\documentclass{revtex4}

\usepackage[latin1]{inputenc}
\usepackage[english]{babel}
\usepackage{graphicx}
\usepackage{amsfonts,amsmath,amssymb}

\begin{document}

\title{Non equilibrium steady states: fluctuations and large deviations of the
density and of the current }

\author{Bernard Derrida}\email{bernard.derrida@lps.ens.fr}
\affiliation{Laboratoire de Physique Statistique,\\ \'Ecole
Normale Sup\'erieure,\\24, rue Lhomond, 75231 Paris Cedex 05,
France}

\date{\today}

\begin{abstract}
These lecture notes give  a short review of methods such as the matrix ansatz, the additivity principle or the macroscopic fluctuation theory,  developed recently in the theory of non-equilibrium phenomena. They show how these methods allow to calculate   the fluctuations and large deviations of the density and of the current  in non-equilibrium steady states of systems like exclusion processes.
The properties of these fluctuations and large deviation functions in non-equilibrium steady states  (for example non-Gaussian fluctuations of density or non-convexity of the large deviation function which generalizes the notion of free energy)  are compared  with  those  of  systems at equilibrium.
\end{abstract}

\maketitle

\section{Introduction}
\label{intro}

The goal of these lectures, delivered at the Newton Institute in Cambridge
for the workshop "Non-Equilibrium Dynamics of Interacting Particle
Systems" in March-April 
2006, is to try to introduce  some methods used to study non-equilibrium steady
 states for systems with stochastic dynamics and  to review some results
 obtained recently on the fluctuations and the large deviations of the density and
 of the current for such systems.
\\ \ \\
Let us start with a few examples of non-equilibrium steady states:
\begin{enumerate}
\item 
{\it A system in contact with two heat baths at temperatures $T_a$ and $T_b$.}
\begin{figure}[ht]
\centerline{\includegraphics[width=8cm]{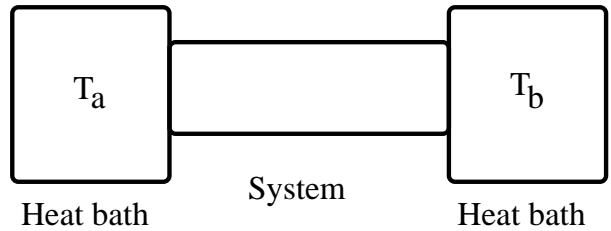}}
\caption{A system in contact with two heat baths at temperatures $T_a$ and $T_b$.}
\label{heatbath}
\end{figure}

At equilibrium, i.e. when the two heat baths are at the same temperature
($T_a=T_b=T$),
the probability $P( C)$ of finding the system in a certain
microscopic configuration $C$ is given by the usual Boltzmann-Gibbs weight
\begin{equation}
P_\text{equilibrium}( C) = Z^{-1} \exp \left[ - {E( C) \over k T} \right]
\label{Boltzmann}
\end{equation}
where $E(C)$ is the internal energy of the system in configuration $C$.
Then the  task of equilibrium statistical mechanics is to derive macroscopic properties (equations of states, phase diagrams, fluctuations,...) from (\ref{Boltzmann}) as a starting point.
A very simplifying aspect of (\ref{Boltzmann}) is  that it  depends neither  on
the precise nature of the
couplings with the heat baths (at least when these couplings are weak) nor on the details of the  dynamics.

When
 the two temperatures $T_a$ and $T_b$ are different, the system reaches
in the long time limit a non-equilibrium steady state \cite{BLR,LLP,EPR1,ST}, but  there
does not exist 
\cite{ruelle1,ruelle2}  
an expression  which generalizes (\ref{Boltzmann}) for the
steady state weights $P( C)$ of the microscopic configurations. 
$$P_\text{non-equilibrium}( C) = \ ? $$
In fact for a non-equilibrium system, the steady state measure $P(C)$
depends in general on the  dynamics of
the system and on its couplings with the heat baths.

Beyond trying to know the steady state measure $P( C)$, which can be done
only for very few examples \cite{KLS,A,DDM,DEHP,SD,DJLS,EMZ}, one might wish to   determine a number of  properties of  non-equilibrium steady states like 
the temperature or energy profiles \cite{EY1,EY2}, the average flow of
energy  through the system \cite{BLL,SL,NR,LLP1}, the probability distribution of this energy flow,  the fluctuations of the internal energy or of the density.

\item {\it A system in contact with two reservoirs of particles at
densities $\rho_a$ and $\rho_b$.} 

Another non-equilibrium steady state situation one can consider is that of a system exchanging particles with two reservoirs \cite{ELS} at densities $\rho_a$ and $\rho_b$.
When $\rho_a \neq \rho_b$ (and in absence of  external field) there is a flow of particles through the system. One can then ask the same questions as for the previous case: for example what is the average current of particles between the two reservoirs, what is the density profile through the system, what are the fluctuations or the large deviations of this current or  of the density.

\begin{figure}[ht]
\centerline{\includegraphics[width=8cm]{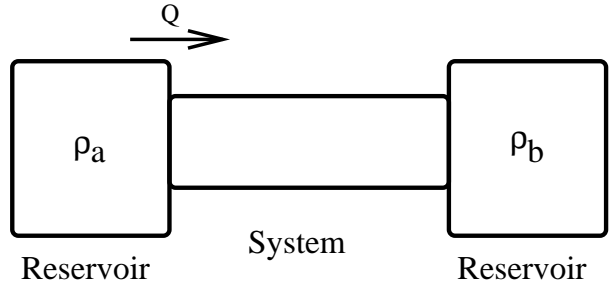}}
\caption{A system in contact with two reservoirs at densities $\rho_a$ and $\rho_b$.}
\label{reservoir}
\end{figure}

\item {\it The symmetric simple exclusion process (SSEP)}

\begin{figure}[ht]
\centerline{\includegraphics[width=10cm]{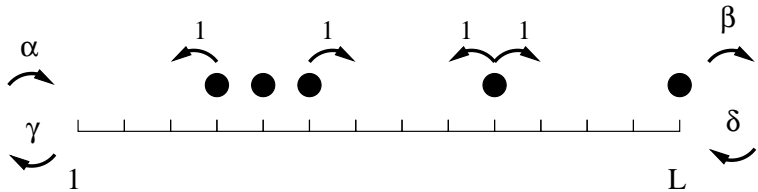}}
\caption{The symmetric simple exclusion process. }
\label{ssep}
\end{figure}

The SSEP    \cite{KOV,HS,Liggett,KL}   is one of the simplest models of a system maintained out of equilibrium by contact with two reservoirs at densities $\rho_a$ and $\rho_b$.
The model is defined as a one-dimensional lattice of $L$ sites with open boundaries,
 each site being either
occupied by a single particle or empty.  During every infinitesimal time
interval $dt$, each particle has a probability $dt$ of jumping to its left
 neighboring site if this site is  empty, and a probability $dt$ of jumping to
its 
right  neighboring site if this  right neighboring site is empty. At the two
boundaries the dynamics is modified to mimic the coupling with reservoirs
of particles: at the left boundary, during each time interval $dt$, a
particle is injected on site $1$ with probability $\alpha dt$ (if this
site is empty) and a particle is removed from site $1$ with probability
$\gamma dt$ (if this site is occupied). Similarly on site $L$, particles
are injected at rate $\delta$ and  removed at  rate $\beta$.

We will see ((\ref{average-profile}) below and \cite{DLS1,DLS2,ED})   that these choices of the rates $\alpha,\gamma,\beta,\delta$ correspond to
the left boundary being connected to a reservoir at density $\rho_a$ and the right boundary to a reservoir at density 
$\rho_b$ with $\rho_a$ and $\rho_b$ given by
\begin{equation}
\rho_a = {\alpha \over \alpha + \gamma} \ \ \ \ ; \ \ \ \ \rho_b = {\delta \over \beta + \delta} \; .
\label{rhoarhobdef}
\end{equation}

One can also think of the SSEP as a simple model of heat transport, if one interprets the particles as quanta of energy. Then if each particle carries  an energy $\epsilon$, the SSEP becomes the model of a system in contact with two heat baths at temperatures $T_a$ and $T_b$  given by (see next section)
\begin{equation}
\exp \left[ {\epsilon \over k T_a}\right] = {\alpha \over  \gamma} \ \ \ \ ; \ \ \ \ 
\exp \left[ {\epsilon \over k T_b}\right] = {\delta \over  \beta}  \; .
\label{TaTbdef}
\end{equation}

\item {\it  Driven diffusive systems}

One can add to the  systems described above  an electric or a gravity field 
which tends to push the particles  in  a preferred  direction.

For example adding   a field to the SSEP means
 that the hopping rates to the left  become  $q$ (the hopping rates to
the right  being still 1). The model becomes then the ASEP (the
asymmetric simple exclusion process) \cite{ABL,FKS,F,K,DDM}  which appears in
many contexts \cite{B,SZL}, such as hopping conductivity \cite{Richards},
models of traffic \cite{CSS}, growth \cite{HHZ}  or polymer dynamics \cite{WVD}. 
\begin{figure}
\centerline{\includegraphics[width=10cm]{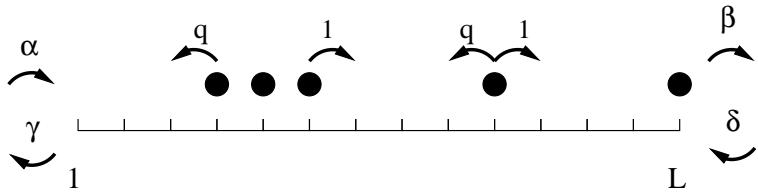}}
\caption{The asymmetric simple exclusion process. }
\label{asep}
\end{figure}
In presence of this external field, the system reaches a non-equilibrium steady state even for a ring geometry, without  need of a reservoir.

The large scale of the ASEP differs noticeably from the SSEP. For example
in the ASEP on the infinite line, one can observe shock  waves whereas
the SSEP is purely diffusive. In fact on large scales  the ASEP is
described \cite{HHZ} by the Kardar Parisi Zhang equation \cite{KPZ} while the SSEP is in the
universalily class of the Edwards Wilkinson equation \cite{SS,EW}.
\end{enumerate}
\ \\ \ \\ 
The outline of these lectures is  as follows:

In section \ref{gen-det-bal} it is recalled how detailed balance should be modified to describe systems in contact with several heat baths at unequal temperatures or several reservoirs at different densities.

In section \ref{free-energy}
 the  large deviation functional of the density is introduced and there is a comparison between its properties in equilibrium and in non-equilibrium steady states.

In section  \ref{non-loc}, the connexion between the non-locality of
the large deviation functional of the density and the presence of long range correlations is discussed.

In section \ref{ssep-sec}  it is shown how to write the evolution
equations of the profile and of the correlation functions for the
symmetric simple exclusion process.

Section \ref{matrix-ssep} describes the matrix ansatz \cite{DEHP} which gives an exact expression of the weights in the non-equilibrium steady state of the symmetric exclusion process.

Using an additivity relation   \ref{add-ssep} established as a
consequence of the matrix ansatz, the large deviation functional
\cite{DLS1,DLS2} of the density for the SSEP is calculated in
\ref{ld-dens-ssep}.

The macroscopic fluctuation theory of Bertini,
  De Sole,   Gabrielli, 
Jona-Lasinio and   Landim \cite{BDGJL1,BDGJL2,BDGJL3,BDGJL4}
  is recalled in section
\ref{mft}, which shows how the calculation of large deviation functional of the density can be formulated as an optimisation problem.

The definition of the large deviation function of the current and the
fluctuation theorem \cite{ECM,GC,ES,Kurchan,LS}  are recalled in section
\ref{ldc} from which  the fluctuation-dissipation theorem for energy or
particle currents  can be recovered 
(section \ref{fdt}).

A perturbative approach \cite{DDR} to calculate the large deviation function of the current
for the SSEP is sketched in
\ref{cur-ssep}.

The additivity principle, which predicts the cumulants and the large deviation function of the current, is presented in section
\ref{add}.

The last four sections are devoted to the ASEP:
the matrix ansatz for the ASEP is recalled in section \ref{matrix-tasep}. It is shown in
section \ref{phas-diag} how to obtain the phase diagram of the TASEP from the matrix ansatz.
An additivity relation from which one can compute the large deviation
funtional of the density \cite{DLS3,DLS4} is established in section
\ref{add-tasep}.
Latsly in section 
\ref{brown-excurs} it is shown that the fluctuations of density are
non-Gaussian  \cite{DEL} in the maximal current phase of the TASEP.

\section{How to generalize detailed balance to non-equilibrium systems }
\label{gen-det-bal}

As in non-equilibrium systems, the steady state measure $P(C)$ depends on the couplings to the heat baths and on the dynamics of the system, each model of a non-equilibrium has to incorporate
a description of  these couplings  and of the dynamics 
(various ways of modeling the effect of heat baths or of reservoirs are described in
see \cite{BLR,EPR2}).
It is often theoretically simpler to represent the effect of the heat baths (or of the reservoirs of particles) by some stochastic terms  such as Langevin forces corresponding to the temperatures of the heat baths.
In practise the dynamics becomes a Markov process.

For a system with stochastic dynamics given by a Markov process (such as the  SSEP or mechanical systems with heat baths represented by Langevin forces)
 the evolution is  specified by a transition matrix $W(C',C)$
which represents the rate at which the system jumps from a configuration
$C$ to  a configuration $C'$ (i.e. the probability that the system jumps
from $C$ to $C'$ during an infinitesimal time interval $dt$ is given by
$W(C',C)dt$). For simplicity, we will limit the discussion to the case
where the total number of accessible configurations is finite. The
probability  $P_t(C)$ of finding the system 
in configuration $C$  at time $t$ evolves therefore according to the  Master equation
\begin{equation}
{d P_t(C) \over dt} =  \sum_{C'} W(C,C') P_t(C') - W(C',C) P_t(C) 
 \ .
\label{markov}
\end{equation}
One can then wonder what should be assumed on the transition matrix $W(C',C)$
to describe a system in contact with one or several heat baths (as for example in figure \ref{heatbath}).
\\ \ \\
At equilibrium, (i.e. when  the system is in contact with a single heat bath at temperature $T$) one usually requires that the transition matrix satisfies {\it detailed balance}
\begin{equation}
W(C',C) \ e^{-{E(C) \over k  T}}
=W(C,C') \ e^{-{E(C') \over k  T}} \ .
\label{detailed-balance}
\end{equation}
This ensures
the time reversal symmetry of
the microscopic dynamics: at equilibrium (i.e. if the initial condition is chosen according to (\ref{Boltzmann})), the probability of observing any given history of the system $\{C_s, 0<s<t\}$ is equal to the probability of observing the reversed history
\begin{equation}
\text{Pro}(\{C_s, 0<s<t\})=
\text{Pro}(\{C_{t-s}, 0<s<t\}) \ .
\end{equation}
Therefore if $\epsilon$ is the energy transferred from the heat bath at
temperature $T$ to the system,  and  $W_\epsilon(C',C)dt$ is the probability that
the system jumps during $dt$ from $C$ to $C'$ by receiving an energy $\epsilon$
from the heat bath, one can rewrite the detailed balance condition (\ref{detailed-balance}) as
\begin{equation}
\ W_\epsilon(C',C)
= e^{-{\epsilon \over k T}} \ 
W_{-\epsilon}(C,C') \ .
\label{detailed-balance-bis}
\end{equation}

If  detailed balance gives a good description of the coupling with a single  heat bath at temperature $T$,  the
straightforward generalization of (\ref{detailed-balance-bis})  
for a system coupled to two heat baths at unequal temperatures like in figure \ref{heatbath} is  \cite{BD4}
\begin{equation}
 W_{\epsilon_a,\epsilon_b}(C',C) =
e^{-{\epsilon_a \over k T_a}- {\epsilon_b \over k T_b} } 
\ 
W_{-\epsilon_a,-\epsilon_b}(C,C')
\label{generalized-detailed-balance}                                         
  \end{equation}
where $\epsilon_a, \epsilon_b$ are the energies transferred from the heat baths at temperatures $T_a, T_b$ to the system when the system jumps from configuration $C$ to configuration $C'$. 
By comparing with (\ref{detailed-balance-bis}), this simply means that
the exchanges of energy with the heat bath at temperature
$T_a$ tend to equilibrate the system at temperature $T_a$  and  the
exchanges with the heat bath at temperature $T_b$ tend to equilibrate
the system at temperature $T_b$.

For a system in contact with two reservoirs of particles at fugacities $z_a$ and $z_b$, as in figure \ref{reservoir},
the generalized detailed balance (\ref{generalized-detailed-balance}) becomes
\begin{equation}
 W_{q_a,q_b}(C',C) 
=
z_a^{q_a} z_b^{q_b} \
W_{-q_a,-q_b}(C,C')
\label{generalized-detailed-balance-bis}            
  \end{equation}
where $q_a$ and $q_b$ are the numbers of particles transferred from the two reservoirs to the system when the system jumps from configuration $C$ to configuration $C'$.
\\ \ \\
From the definition of the dynamics of the SSEP, it is easy to check that it satisfies (\ref{generalized-detailed-balance-bis}) with
\begin{equation}
z_a = {\alpha \over \gamma} \ \ \ ; \ \ \
z_b = {\delta \over \beta} 
\label{za-zb}
\end{equation}
One can also check from (\ref{TaTbdef}) that if one interprets the particles as quanta of energy, (\ref{generalized-detailed-balance}) is satisfied.
\\ \ 

One way of justifying (\ref{generalized-detailed-balance}) is to consider the composite system made up of the system we want to study and of  the two reservoirs. This composite system is isolated and therefore its total energy ${\cal E}$
\begin{equation}
{\cal E}= E(C)+ E_a + E_b
\label{calE}
\end{equation}
 is conserved by the dynamics. 
In (\ref{calE})  $E(C)$ is the energy of the system we want to study and $E_a,E_b.$ are the energies of the two reservoirs (for simplicity we assume that the energy of the coupling between the reservoirs and the system is small).
 Whenever there is an evolution step in the dyanmics, the system jumps from the microcopic configuration $C$ to the configuration $C'$ and the energies of the reservoirs jump from $E_a,E_b$ to $E_a', E_b'$. For the composite system to be able to reach the microcanonical distribution and for microcanonical detailed balance to hold one needs that   the transition rates satisfy 
\begin{equation}
e^{S_a(E_a) + S_b(E_b) \over k} \text{Pro}(\{C,E_a,E_b\} \to \{C', E'_a, E_b'\})= 
e^{S_a(E_a') + S_b(E_b') \over k} \text{Pro}(\{C',E_a',E_b'\} \to \{C, E_a, E_b\})
\label{micro}
\end{equation}
where $S_a(E_a)$ and $S_b(E_b)$ are the  entropies of the two reservoirs at  energies $E_a$ and $E_b$. 
Then if the heat baths are large enough, one has 
\begin{equation}
S(E_a)-S(E_a') = {E_a - E_a' \over T_a} \ \ \ \ ; 
\ \ \ \ S(E_b)-S(E_b') = {E_b - E_b' \over T_b} 
\end{equation}
where $T_a$ and $T_b$ are the (microcanonical) temperatures of the two heat baths and (\ref{micro}) reduces to (\ref{generalized-detailed-balance}).
\ \\ \ \\ 
{\bf Remark:}  
The quantity $- {\epsilon_a \over k T_a}-{ \epsilon_b \over k T_b}$ in (\ref{generalized-detailed-balance}) is the  entropy produced in the reservoirs. 
In fact, in the theory of non-equilibrium phenomena, one can associate to   an arbitrary Markov process, defined by  transition rates $W(C',C)$, 
an entropy production \cite{ECM,GC,ES,LS,MRW,Seifert}  (in the surrounding heat baths) given by
$$\Delta S(C \to C') = \log {W(C',C) \over W(C,C')}$$
and  (\ref{generalized-detailed-balance})  appears as one particular case of this general definition.

\section{Free energy and the large deviation function }
\label{free-energy}
At  equilibrium the free energy is defined by
\[
F= -k T \log Z =  -k T \log \left[ \sum_C \exp \left(-{E(C) \over T} \right) \right] \]
In this section
 we are going to see 
 that the knowledge of  the free energy  gives also the distribution of
the fluctuations and the large deviation function
of the density. This will enable us to 
 extend  the notion of free energy to non-equilibrium systems
by considering the large deviation functional  \cite{KOV,E,DZ}  of the density.

\begin{figure}[ht]
\centerline{\includegraphics[width=8cm]{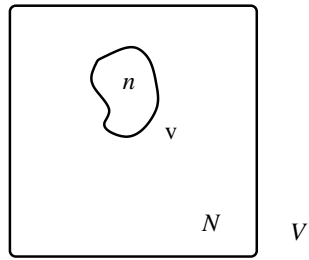}}
\caption{For a system of $N$ particles in total volume $V$,  the
probability $P_v(n)$ of having $n$ particles in a large subvolume $v$ is given  by (\ref{pv}).}
\label{ld-density}
\end{figure}

\begin{figure}[ht]
\centerline{\includegraphics[width=7cm]{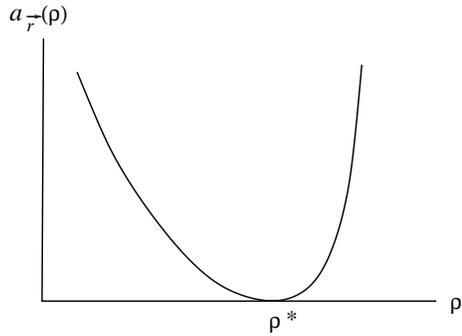}}
\caption{A typical shape of the large deviation function $a_{\vec{r}}(\rho)$. The most likely density $\rho^*$ is the value where  $a_{\vec{r}}(\rho)$ vanishes.}
\label{ar}
\end{figure}

 If one considers a  box  of volume $V$ containing $N$ particles as in figure \ref{ld-density},
the probability $P_v(n)$ of finding $n$ particles in a subvolume $v$
located near a position $\vec{r}$ has the following  large $v$ dependence 
\begin{equation}
P_v(n) \sim \exp \left[  - v \; a_{\vec{r}}\left({n \over v} \right)\right]
\label{pv}
\end{equation}
where $a_{\vec{r}}(\rho)$ is a large deviation function.
Figure \ref{ar} shows a  typical shape of $a_{\vec{r}}(\rho)$ for an homogeneous
system (i.e. not   at a coexistence between different phases) with a
single minimum at $\rho=\rho^*$ where $a_{\vec{r}}(\rho)$ vanishes.

One can also define the large deviation functional ${\cal F}$ for an arbitrary density profile.
If one divides  a system of linear size $L$ into  $n$ boxes of linear
size $l$ (in dimension $d$, one has  of course $n = L^d/l^d$ such boxes), one can try to determine the probability of finding a certain density profile $\{ \rho_1, \rho_2, ... \rho_n\}$, i.e. the probability of seeing $l^d \rho_1$ particles in the first box, $l^d \rho_2$
particles in the second box, ... $l^d \rho_n$ in the $n$th box. For large $L$ one expects the following $L$ dependence of this probability
\begin{figure}[ht]
\centerline{\includegraphics[width=8cm]{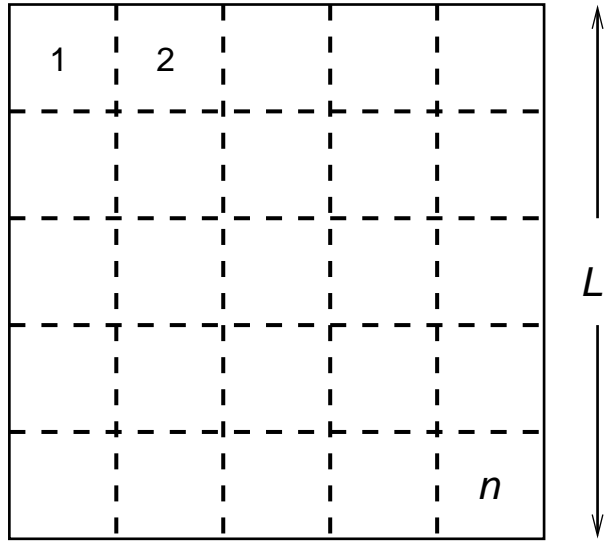}}
\caption{In (\ref{finite}) one specifies the densities $\rho_i$ in each box $i$ }
\label{functional}
\end{figure}
\begin{equation}
{\rm Pro}(\rho_1,... \rho_n) \sim \exp[ - L^d {\cal F}(\rho_1, \rho_2, ...\rho_n)]
\label{finite}
\end{equation}
where ${\cal F}$ is a large deviation function  which generalizes $a_{\vec{r}}(\rho)$ defined in (\ref{pv}).
If one introduces a reduced coordinate $\vec{x}$ 
\begin{equation}
\vec{r}= L \vec{x}
\end{equation}
and if one takes the limit $L \to \infty$, $l \to \infty$ with $l \ll L$
so that the number $n$ of boxes becomes large, this becomes a functional ${\cal F}(\rho(\vec{x}))$ for an arbitrary density profile $\rho(\vec{x})$
\begin{equation}
{\rm Pro}(\rho(\vec{x})) \sim \exp[ - L^d {\cal F}(\rho(\vec{x}))]  \; .
\label{continu}
\end{equation}
Clearly the large deviation function $a_{\vec{r}}(\rho)$ or the large deviation
functional ${\cal F}(\rho(\vec{x}))$  can be defined for equilibrium systems as well as for non-equilibrium systems.

For equilibrium systems, one can show  that $a_{\vec{r}}(\rho)$   is closely related
to the free energy: if the volume $v$ is sufficiently large, 
for short ranged interactions   and in absence of external potential, 
 the large deviation function $a_{\vec{r}}(\rho)$ is independent of $\vec{r}$
and its expression is 
given by
\begin{equation}
a_{\vec{r}}(\rho)=a(\rho)={f(\rho) - f(\rho^*) -  (\rho-\rho^*) f'(\rho^*) \over kT}
\label{areq}
\end{equation}
where $f(\rho)$ is the free energy per unit volume at density $\rho$ and
$\rho^* = {N \over V}$.
This can be seen by noticing that 
if $v^{1/d} $ is much larger than the
range of the interactions
and if  $ v \ll V$  one has
\begin{equation}
P_v(n)= {Z_v(n) \ Z_{V-v}(N-n) \over Z_V(N)} \exp\left[O(v^{d-1 \over d}) \right]
\label{pv1}
\end{equation}
where $Z_V(N)$ is the partition function of $N$ particles in a volume $V$ and the term $ \exp\left[O(v^{d-1 \over d}) \right]$ represents the
interactions between all
pairs of particles, one of which is the volume $v$ and the other one in $V-v$. Then taking the $\log$ of (\ref{pv1}) and using the fact that
the free energy $f(\rho)$ per unit volume is defined by 
\begin{equation} \lim_{V \to \infty} {\log Z_V(V\rho) \over V}= -{f(\rho) \over kT} \label{freeenergy}
\end{equation} one gets   (\ref{areq}).
The functional ${\cal F}$ can also be expressed in terms of 
$f(\rho)$:
if one considers $V \rho^*$  particles in a volume $V=L^d$, one can generalize (\ref{pv1})  
for systems with short range  interactions and no external potential
\begin{equation}
{\rm Pro}(\rho_1,... \rho_n)= {Z_v(v\rho_1) ... Z_v(v\rho_n) \over Z_V(V r)} \exp\left[O\left( {L^d \over l}\right) \right]
\end{equation}
where $v= l^d$.
 Comparing with (\ref{finite}),  in the limit  $L \to \infty, \ l \to \infty$, keeping $n$ fixed gives
\begin{equation}
{\cal F}(\rho_1, \rho_2, ...\rho_n)= {1 \over kT} {1 \over n}
\sum_{i=1}^n [ f(\rho_i) - f(\rho^*)] \; .
\end{equation}
In the limit of an infinite number of boxes,  this becomes
\begin{equation}
 {\cal F}(\rho(\vec{x}))={1 \over k T} \int d \vec{x} \ [
f(\rho(\vec{x})) - f(\rho^*)] \; .
\label{Feq}
\end{equation}
Thus for a system at equilibrium, the large deviation functional ${\cal F}$  is fully determined by the knowledge
of the free energy $f(\rho)$
per unit volume.
In (\ref{Feq}), we see that 
\begin{itemize}
\item
The functional ${\cal F }$ is a {\it local} functional
of $\rho(\vec{x})$. 
\item
 It is  also a {\it convex} functional of the profile $\rho(\vec{x})$ (as the free energy $f(\rho)$ is a
convex function of  the density $\rho$, i.e. $f(\alpha \rho_1 + (1- \alpha) \rho_2) \leq \alpha f(\rho_1) + (1- \alpha) f(\rho_2)$ for $0<\alpha <1$) . 
\item 
When $f(\rho)$ can be expanded around $\rho^*$ (i.e. at densities where the free energy $f(\rho)$ is not singular)
 one obtains also from (\ref{Feq}) that the fluctuations
 of the density profile are Gaussian. In fact if one expands (\ref{areq}) near $\rho^*$  and one replaces it into (\ref{pv}) one gets
that the distribution of the number $n$ of particles in the subvolume $v$ is Gaussian (if $v$ is large enough)
\begin{equation}
P_v(n) \sim \exp \left[  - \;  v {f''(\rho^*) \over 2 k T} (\rho - \rho^*)^2 \right]
= \exp \left[  - \;   {f''(\rho^*)\over  2  \; v  \; k T} (n  -  v \rho^*)^2 \right]
\label{pvbis}
\end{equation}
and its variance, as predicted by Smoluchowki and Einstein, is given by
\begin{equation}
\langle n^2 \rangle - \langle n \rangle^2 =  v {k T  \over  f''(\rho^*)} = v  \; k  T \; \kappa(\rho^*)
\end{equation}
where the compressibility $\kappa(\rho)$ is defined by
\begin{equation}
\kappa(\rho) = {1 \over \rho}{d \rho \over dp}
\label{compressibility}
\end{equation}
(and  the pressure $p$ is  given as usual by $p = - {d \over dV} V f({N \over V}) = \rho^* f'(\rho^*) - f(\rho^*)$).

Note that at  a phase transition, $f(\rho)$ is singular and the fluctuations of density are in general non-Gaussian.
\item 
One also knows (by the Landau argument) that, with short range
 interactions, there is no phase transition if the dimension of space is  one dimension.
\end{itemize}
In contrast to equilibrium systems, one can observe in non-equilibrium steady states of systems such as the ones described in figures \ref{heatbath} and \ref{reservoir} 
\begin{itemize}
\item
The large deviation functional $\cal F$ may be non local. For example in the case of the SSEP, we will see in section \ref{ld-dens-ssep} that the functional is given by for $\rho_a-\rho_b$ small by (see (\ref{rhoa-rhob-small})  below).
\begin{align}
{\cal F}(\{\rho(x)\}) = & \int_0^1 dx  \left[ \rho(x) \log {\rho(x) \over \rho^*(x)} + (1-\rho(x)) \log {1- \rho(x) \over 1 - \rho^*(x)} \right]
\label{secondorder}
\\ \nonumber
& +  {(\rho_a - \rho_b)^2 \over [\rho_a (1- \rho_a)]^2}
\int_0^1 dx \int_x^1 dy\: x(1-y)
  \big(\rho(x)-\rho^*(x)\big)
  \big(\rho(y)-\rho^*(y)\big)
 + O(\rho_a-\rho_b)^3  
\end{align}
where
  $\rho^*(x)$ is 
the most likely profile 
\begin{equation}
 \rho^*(x) = (1-x)  \rho_a + x \rho_b \ .
\label{rhostar}
\end{equation} 
\item
For the ASEP, there is a range of parameters where the  functional $\cal F$ is  non convex (see \cite{DLS3,DLS4} and section \ref{add-tasep} below).
\item
There are also cases, where  in the maximal  current phase, the density   fluctuations are non Gaussian (see \cite{DEL} and section \ref{brown-excurs} below).
\item
In non-equilibrium systems nothing prevents the existence of phase
transitions in one dimension
\cite{K,DDM,DEHP,SD,EFGM,Mal,EKKM,KSKS,Der,PS,Evans,KLMST,PFF,CDE,FF}.

\end{itemize}

\section{Non locality of the  large deviation functional of the density and long range correlations }
\label{non-loc}
A feature characteristic of non-equilibrium systems is the presence of
weak long range correlations  \cite{Spohn,SC,DKS,OS,DELO,DLS5}. For example for the SSEP, we will see \cite{Spohn} in next section  (\ref{corr},\ref{corr-bis})   that for large $L$ the correlation function of the density is given for $0<x<y<1$ 
\begin{equation}
\langle \rho(x) \rho(y) \rangle_c = - {(\rho_a-\rho_b)^2 \over L} x (1-y) 
\label{2pt-function}
\end{equation}
The presence of these long-range correlations is directly related to the non-locality of the large deviation functional $\cal F$. This can be seen by introducing the generating function ${\cal G}(\{\alpha(x)\})$ of the density defined by
\begin{equation}
\exp \left[L {\cal G}(\{\alpha(x)\}) \right] = \left\langle 
\exp \left[L \int_0^1 \alpha(x) \rho(x) dx  \right]  \right\rangle 
\label{Gdef}
\end{equation}
where $\langle . \rangle$ is an average over the  profile  $\rho(x)$ in the
steady state. As the probability of this profile  is given by
(\ref{continu}) the  average  in (\ref{Gdef}) is dominated, for large $L$,  by an optimal profile, which depends on $\alpha(x)$, and $\cal G$ is the Legendre transform of $\cal F$
\begin{equation}
{\cal G}(\{\alpha(x)\})  = \max_{\{\rho(x)\}}  \left[ \int_0^1 \alpha(x)
\rho(x) dx   -{\cal F} (\{\rho(x)\}) \right]   \ . 
\label{legendre}
\end{equation}
It is clear from (\ref{legendre}) that if the large deviation ${\cal F}$ is local (as in (\ref{Feq})), then the generating function $\cal G$ is also local. Now by taking derivatives with respect to $\alpha(x)$ one gets that the average profile and the correlation functions are given  by
\begin{equation}
\rho^*(x) \equiv  \langle \rho(x) \rangle=  \left. {\delta {\cal G} \over \delta \alpha(x)}\right|_{\alpha(x)=0} 
\label{1pt}
\end{equation}
\begin{equation}
\langle \rho(x)  \rho(y) \rangle_c \equiv
\langle \rho(x)  \rho(y) \rangle
-\langle \rho(x) \rangle \langle  \rho(y) \rangle
= {1 \over L} \left. {\delta^2 {\cal G} \over \delta \alpha(x) \;  \delta \alpha(y)}\right|_{\alpha(x)=0} 
\label{2pt}
\end{equation}
This shows that the non-loacality of $\cal G$ is directly related to the existence of long range correlations.

One can understand  the $L$ dependence in   (\ref{2pt})  by assuming
that the non-local functional $\cal G$ is can be expanded as
\begin{equation}
{\cal G}(\alpha(x)) = 
\int_0^1 dx \  A(x)  \ \alpha(x)  + 
\int_0^1 dx  \ B(x)  \ \alpha(x)^2  + 
\int_0^1 dx \int_x^1 dy \  C(x,y) \ \alpha(x) \  \alpha(y) + ...
\label{continuous}
\end{equation}
If one comes back to a discrete system of $L$ sites with a number $n_i$ of particles on site $i$, one has
\begin{equation}
L {\cal G}(\alpha(x))  \simeq \log \left[ \left\langle  \exp\sum_i \alpha_i n_i\right\rangle \right]
\end{equation}
By expanding in powers of the $\alpha_i$ one has
\begin{equation}
L {\cal G}(\alpha(x))  \simeq  \sum_{i=1}^L A_i \alpha_i +
  \sum_{i=1}^L B_i \alpha_i^2 +
  \sum_{i<j} C_{i,j} \alpha_i \alpha_j + ...
\label{discrete}
\end{equation}
Clearly one has
\begin{equation}
\langle n_i \rangle = A_i \ \ \ ; 
 \ \ \  
\langle n_i ^2\rangle_c = 2 B_i \ \ \ ; 
 \ \ \  
\langle n_i  n_j \rangle_c = C_{i,j} \ \ \  
\end{equation}
and comparing (\ref{continuous}) and (\ref{discrete}) one sees that
\begin{equation}
C_{i,j} = {1 \over L}C\left( {i \over L} , {j \over L} \right)
\end{equation}
which leads to (\ref{2pt}). A similar reasoning would show that
\begin{equation}
\langle \rho(x_1)  \rho(x_2)... \rho(x_k)  \rangle_c 
= {1 \over L^{k-1}} \left. {\delta^k {\cal G} \over \delta \alpha(x_1) ... \delta \alpha(x_k) }\right|_{\alpha(x)=0}  \ .
\label{kpt}
\end{equation}
This $1/L^{k-1}$  dependence of the $k$ point function can indeed be proved in the SSEP \cite{DLS5}.
We see that  all the correlation functions   can in principle be obtained 
by expanding,  when this expansion is meaningful (see \cite{DLS3,DLS4} for counter-examples), 
 the large deviation function $\cal G$ in
powers of  $\alpha(x)$.

\section{The  symmetric simple exclusion model}
\label{ssep-sec}

For the SSEP, the calculation of the average profile or of the correlation functions can be done directly from 
 the  definition of the model.  If
$\tau_i=0$ or $1$ is a binary variable  indicating whether site $i$ is
occupied or empty, one can  write the time evolution of the average occupation $\langle
\tau_i \rangle$
\begin{align}
{d \langle \tau_1 \rangle \over dt } = & \alpha - (\alpha + \gamma + 1 )
\langle \tau_1 \rangle + \langle \tau_2 \rangle
\nonumber \\
{d \langle \tau_i \rangle \over dt } = &
\langle \tau_{i-1} \rangle -2  \langle \tau_i \rangle +  \langle
\tau_{i+1} \rangle                  \ \ \ \  \ \ {\rm for} \ \  2 \leq i
\leq L-1                     \label{evolution}                                     \\
{d \langle \tau_L \rangle \over dt } = &
\langle \tau_{L-1} \rangle -(1+ \beta + \delta)  \langle \tau_L \rangle +
\delta   \nonumber                 \end{align}
The steady state density profile (obtained by writing that ${d
\langle \tau_i \rangle \over dt } =0$) is \cite{DLS2}
\begin{equation}
\langle \tau_i \rangle = {\rho_a (L+ {1 \over \beta + \delta}-i) + \rho_b
(i-1 +{1 \over \alpha + \gamma}) \over L + {1 \over \alpha + \gamma}+{1
\over \beta + \delta}-1} \; \label{profile}
\end{equation}
with $\rho_a$ and $\rho_b$  defined as in (\ref{rhoarhobdef}).
One can notice that for large $L$, if one introduces a macroscopic coordinate $i = Lx$, this becomes
\begin{equation}
\langle \tau_i \rangle = \rho^*(x) = (1-x) \rho_a + x \rho_b
\label{average-profile}
\end{equation}
and one recovers (\ref{rhostar}).
For large $L$
one can also remark that 
 $\langle \tau_1 \rangle \to \rho_a$ and $\langle \tau_L \rangle \to
\rho_b$  indicating that  $\rho_a$ and $\rho_b$ defined by  (\ref{rhoarhobdef}) represent the densities of the left and right reservoirs.
One can in fact show \cite{DLS1,DLS2,ED} that the rates
$\alpha,\gamma,\beta,\delta$ do correspond to
the left and right boundaries being connected respectively  to  reservoirs
at densities $\rho_a$
and
$\rho_b$.

The average current in the steady state  is given by
\begin{equation}
\langle J \rangle =
\langle \tau_i (1-\tau_{i+1}) - \tau_{i+1}(1- \tau_i) \rangle=
\langle \tau_i - \tau_{i+1} \rangle=
{\rho_a - \rho_b  \over L + {1 \over \alpha + \gamma}+{1 \over \beta +
\delta}-1}
\label{current}
\end{equation}
This shows that
for large $L$, the current $ \langle J \rangle \simeq {\rho_a - \rho_b \over L }
$
is  proportional to the gradient of the density (with a coefficient of
proportionality which is here simply 1)
and therefore follows Fick's law.

One can  write down the equations which generalize
(\ref{evolution}) and   govern the time
evolution of the two-point function or higher correlations. For example
one finds \cite{Spohn,DLS5} in the steady state for $1 \leq i < j \leq L$
\begin{equation}
\langle\tau_i\tau_i\rangle_c \equiv
\langle\tau_i\tau_j\rangle
- \langle\tau_i\rangle
\langle\tau_j\rangle
 = -
   \frac{({1 \over \alpha + \gamma}+i-1)({1 \over \beta + \delta}+L-j)}
{({1 \over \alpha + \gamma}+{1 \over \beta + \delta}+L-1)^2({1 \over \alpha +
\gamma}+{1 \over \beta + \delta}+L-2)}\:(\rho_a-\rho_b)^2 \ .
\label{corr}
\end{equation}
For large $L$, if one introduces macroscopic coordinates
$i=Lx$ and $j=Ly$, this becomes for $x<y$
\begin{equation}
 \langle\tau_{Lx}\tau_{Ly}\rangle_c= - \frac{x(1-y)}{L} (\rho_a-\rho_b)^2 
\label{corr-bis}
\end{equation}
which is the expression (\ref{2pt-function}).

 One could believe that 
these weak, but long range, correlations 
 play no role in the large $L$ limit. However if one considers macroscopic quantities such as the total number $N$ of particles in the system, one can see that
these two point correlations give a leading contribution to the variance of $N$
\begin{equation}
\langle N^2 \rangle - \langle N \rangle^2 = \sum_i [ \langle \tau_i \rangle - \langle \tau_i \rangle^2 ] + 2 \sum_{i<j} \langle \tau_i \tau_j \rangle_c \simeq L \left[ \int dx \rho^*(x) (1- \rho^*(x) ) - 2 (\rho_a-\rho_b)^2
\int_0^1 dx \int_x^1 dy \    x(1-y) \right]
\end{equation}

For the SSEP, one can write down the steady state equations  satisfied by higher
correlation functions  to get  for example for $x<y<z$
\begin{equation}
 \langle\tau_{Lx}\tau_{Ly} \tau_{Lz}\rangle_c= - 2 \frac{x(1- 2 y)(1-z)}{L^2}
(\rho_a-\rho_b)^3 
\end{equation}
 but solving these equations become quickly
quickly too complicated. We will see in  next section that the matrix ansatz gives an algebraic procedure to 
 calculate all these correlation functions \cite{DLS5}.

\section{The matrix ansatz for the  symmetric exclusion process}
\label{matrix-ssep}
The matrix ansatz is an approach inspired by the construction of exact
eigenstates in quantum spin chains \cite{HN,AKLT,KSZ}. It  gives an algebraic way of calculating exactly the weights of
all the configurations in the steady state.
In \cite{DEHP} it was shown that the probability of  a microscopic
configuration
$\{ \tau_1, \tau_2, ...\tau_L\}$ can be written as the matrix element of
a product of $L$ matrices
\begin{equation}
{\rm Pro}(\{ \tau_1, \tau_2, ...\tau_L\}) = {\langle W | X_1 X_2 ... X_L | V
\rangle \over \langle W | (D+E)^L |V \rangle }
\label{matrix}
\end{equation}
where the matrix $X_i$ depends on the occupation  $\tau_i$ of site $i$
\begin{equation}
X_i =  \tau_i  D + (1 - \tau_i) E
\end{equation}
 and the matrices $D$ and $E$ satisfy the following algebraic rules
\begin{eqnarray}
&& DE-ED= D+E \nonumber \\
&& \langle W | ( \alpha E - \gamma D) = \langle W| \label{algebra} \\
&&  ( \beta D - \delta E)| V \rangle = | V \rangle  \; . \nonumber
\end{eqnarray}

\begin{figure}[ht]
\centerline{\includegraphics[width=8cm]{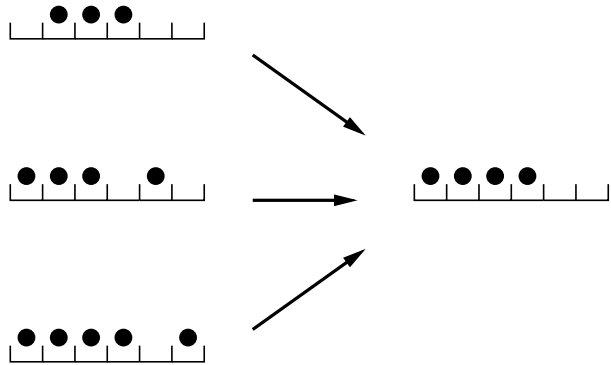}}
\caption{The three configurations which appear on the left hand side of (\ref{conf1}) and from which one can jump to the configuration which appears on the right hand side of (\ref{conf1}). }
\label{proof}
\end{figure}

Let us check on the  simple example  of figure \ref{proof} that  expression (\ref{matrix}) does give the steady state weights: if one chooses the configuration where the first $p$ sites on the left are occupied and the remaining $L-p$ sites on the right are empty, the weight of this configuration is given by
\begin{equation}
{\langle W |D^p E^{L-p} | V
\rangle \over \langle W | (D+E)^L |V \rangle } \ .
\label{conf}
\end{equation}
For (\ref{matrix}) to be the weights of all configurations in the steady
steady, one needs that the rate at which the system enters each configuration and
the rate at which the system leaves it should be equal.
In the case of  the configuration whose weight is (\ref{conf}), this means
that the following steady state identity should be satisfied (see figure
\ref{proof}):
\begin{align}
\alpha {\langle W |E D^{p-1} E^{L-p} | V
\rangle \over \langle W | (D+E)^L |V \rangle }
+{\langle W |D^{p-1}E D  E^{L-p-1} | V
\rangle \over \langle W | (D+E)^L |V \rangle }
+ \beta {\langle W |D^p E^{L-p-1}D | V
\rangle \over \langle W | (D+E)^L |V \rangle }
=
 (\gamma + 1 + \delta) {\langle W |D^p E^{L-p} | V
\rangle \over \langle W | (D+E)^L |V \rangle }
\label{conf1}
\end{align}
 This equality is easy to check by rewriting (\ref{conf1}) as
\begin{align}
 {\langle W |(\alpha E- \gamma D) D^{p-1} E^{L-p} | V
\rangle \over \langle W | (D+E)^L |V \rangle }
-{\langle W |D^{p-1}(DE-E D)  E^{L-p-1} | V
\rangle \over \langle W | (D+E)^L |V \rangle }
+  {\langle W |D^p E^{L-p-1}(\beta D - \delta E) | V
\rangle \over \langle W | (D+E)^L |V \rangle }=0
\end{align}
and by using (\ref{algebra}). A similar reasoning \cite{DEHP}  allows one to prove
that the corresponding steady state identity holds for any other configuration.

A priori one should construct the matrices $D$ and $E$ (which might be
infinite-dimensional \cite{DEHP}) and the vectors
$\langle W|$ and $|V \rangle$ satisfying (\ref{algebra}) to calculate the
weights (\ref{matrix}) of the microscopic configurations.
However these weights do not depend on the particular representation
chosen and can be calculated directly from (\ref{algebra}).
This can be easily seen by using the two matrices $A$ and $B$ defined by
\begin{align}
& A=\beta D - \delta E
\nonumber \\
& B=\alpha  E  - \gamma D
\label{ABdef}
\end{align}
which satisfy
\begin{equation}
 AB - BA = (\alpha \beta - \gamma \delta) (D+E) = (\alpha + \delta)A +
(\beta + \gamma) B  \ .
\label{ABmBA}
\end{equation}
Each product of $D$'s and $E$'s can be written as a sum of products
of $A$'s and $B$'s which can be ordered using
(\ref{ABmBA}) by pushing all the $A$'s to the right and all the $B$'s to
the left. One then gets  a sum of terms of the form $B^p A^q$, the
matrix elements of which can be evaluated easily ($\langle W | B^p A^q | V \rangle
= \langle W | V \rangle $) from (\ref{algebra}) and (\ref{ABdef}).

One can  calculate with the weights (\ref{matrix})  the average density profile
$$ \langle \tau_i \rangle = { \langle W | (D+E)^{i-1} D (D+E)^{L-i} | V
\rangle \over \langle W | (D+E)^{L} | V
\rangle }$$
as well as all the correlation functions 
$$ \langle \tau_i  \tau_j\rangle = { \langle W | (D+E)^{i-1} D
(D+E)^{j-i-1} D (D+E)^{L-j} | V
\rangle \over \langle W | (D+E)^{L} | V
\rangle }$$
and one can recover that way
(\ref{profile}) and (\ref{corr}).

Using the fact that the average current  between sites $i$ and
$i+1$ is given by
$$
\langle J \rangle =
{ \langle W | (D+E)^{i-1} (DE - ED) (D+E)^{L-i-1} | V
\rangle \over \langle W | (D+E)^{L} | V
\rangle } = { \langle W | (D+E)^{L-1} | V
\rangle \over \langle W | (D+E)^{L} | V
\rangle }
$$
(of course in the steady state the current does not depend on $i$)
and from the expression (\ref{current}) one can calculate the normalization
\begin{equation}
\frac{\langle W|(D+E)^L|V\rangle}
       {\langle W|        V\rangle}=
  \frac{1}{(\rho_a-\rho_b)^L}\:
  \frac{\Gamma(L+ {1 \over \alpha + \gamma }+{1\over \beta + \delta})}{\Gamma({1 \over \alpha +
\gamma }+{1\over \beta + \delta}) }
\label{norm}
\end{equation}
(see equation (3.11) of \cite{DLS2} for an alternative derivation of this
expression).
\\ \\ {\bf Remark:} when $\rho_a=\rho_b=r$, the two reservoirs are at
the same density and the steady state becomes the equilibrium (Gibbs)
state of the lattice gas at this density $r   $.
In this case, the weights of the configurations are those of a Bernoulli
measure at density $r   $, that is
\begin{equation}
{\rm Pro}(\{\tau_1, \tau_2, ...\tau_L\}) =  \prod_{i=1}^L [ \; r  \;  \tau_i + (1-
r)(1-\tau_i) ] \  .
\label{equilibrium}
\end{equation}
This  case corresponds to  a limit where the matrices $D$ and $E$
commute  (it can be recovered by making all the calculations with the matrices (\ref{matrix},\ref{algebra}) for $\rho_a \neq
\rho_b$ and
by taking the limit $\rho_a \to \rho_b$ in the final expressions, as all
the expectations, for a lattice of finite size $L$, are rational functions
of $\rho_a$ and $\rho_b$).

 \section{Additivity as a consequence of the matrix ansatz}
\label{add-ssep}
As in (\ref{matrix})  the weight of each configuration is written as
the matrix element of a product of $L$ matrices, one can try to insert at
a position $L_1$ a complete basis in order to relate the properties of a
lattice of $L$ sites to those of two subsystems of sizes $L_1$ and
$L-L_1$.
To do so let us define
 the following left and right eigenvectors of the
operators  \\ $\rho_a E - (1-\rho_a)D$ and
$(1-\rho_b)D-\rho_bE$
\begin{eqnarray}
 && \langle \rho_a,a| \ [\rho_a E-(1-\rho_a)D]=  a \langle  \rho_a,a|
\nonumber \\
 &&  [ (1-\rho_b ) D- \rho_b E ]  \ | \rho_b ,b \rangle = b | \rho_b ,b
\rangle \; . \label{eigen}
\end{eqnarray}
It is easy to see, using the definition (\ref{rhoarhobdef}),  that the
vectors $\langle W |$ and $| V \rangle$
 are given by
\begin{eqnarray}
\langle W | = \langle \rho_a,(\alpha + \gamma)^{-1}| \nonumber \\
 | V \rangle= |\rho_b, (\beta + \delta)^{-1}\rangle \; .
\end{eqnarray}
It is then  possible to show, using simply the fact (\ref{algebra}) that
$DE-ED=D+E$ and
the definition of the eigenvectors (\ref{eigen}),  that (for $\rho_b <
\rho_a$)
\begin{equation}
  \frac{\langle \rho_a,a|Y_1 Y_2|\rho_b,b\rangle}
       {\langle \rho_a,a|       \rho_b,b\rangle}
  =  
 \oint\displaylimits_{\rho_b<|\rho|<\rho_a} \frac{d\rho}{2i\pi}\:
    \frac{(\rho_a-\rho_b)^{a+b}}{(\rho_a-\rho)^{a+b}(\rho-\rho_b)} \:
    \frac{\langle \rho_a,a|Y_1|\rho, b\rangle}
         {\langle \rho_a,a|    \rho, b\rangle} \:
    \frac{\langle \rho,1-b|Y_2|\rho_b,b\rangle}
         {\langle \rho,1-b|    \rho_b,b\rangle}
\label{fermeture}
\end{equation}
where $Y_1$ and $Y_2$ are arbitrary polynomials of the matrices $D$ and $E$.
 \\ \ \\
{\bf Proof of (\ref{fermeture}):} to 
 prove  (\ref{fermeture}) it is sufficient to choose   $Y_1$ of
the form $[\rho_a E-(1-\rho_a)D]^n [D+E]^{n'}$ and $Y_2$ of the form
$[D+E]^{n''} [ (1-\rho_b ) D- \rho_b E ]^{n'''}$  (one can show, using
$DE-ED=D+E$, that any polynomial $Y_1$ or $Y_2$ can be reduced to a finite
sum of such terms).
Then  proving (\ref{fermeture}) for such choices of $Y_1$ and $Y_2$  reduces to proving
 \begin{equation}
  \frac{\langle \rho_a,a|(D+E)^{n'+n''}|\rho_b,b\rangle}
        {\langle \rho_a,a|       \rho_b,b\rangle}
   = 
  \oint\displaylimits_{\rho_b<|\rho|<\rho_a} \frac{d\rho}{2i\pi}\:
     \frac{(\rho_a-\rho_b)^{a+b}}{(\rho_a-\rho)^{a+b}(\rho-\rho_b)} \:
     \frac{\langle \rho_a,a|(D+E)^{n'}|\rho, b\rangle}
          {\langle \rho_a,a|    \rho, b\rangle} \:
     \frac{\langle \rho,1-b|(D+E)^{n''}|\rho_b,b\rangle}
          {\langle \rho,1-b|    \rho_b,b\rangle}
\  .
 \label{fermeture-bis}
 \end{equation}
As from (\ref{norm}) one has
\begin{equation}
\frac{\langle \rho_a,a|(D+E)^{L}|\rho, b\rangle}
         {\langle \rho_a,a|    \rho, b\rangle}  = \frac{\Gamma(L+a+b) }{(\rho_a- \rho_b)^L \  \Gamma(a+b)} \  .
\end{equation}
Then  (\ref{fermeture}) and (\ref{fermeture-bis}) follow as one can easily check that
\begin{equation}
  \frac{ \Gamma(n'+n''+a+b) }
       {(\rho_a-       \rho_b)^{n'+n''}}
  = 
 \oint\displaylimits_{\rho_b<|\rho|<\rho_a} \frac{d\rho}{2i\pi}\:
    \frac{(\rho_a-\rho_b)^{a+b}}{(\rho_a-\rho)^{a+b+n'}(\rho-\rho_b)^{n''+1}} \:
    \frac{\Gamma(n'+a+b) \Gamma(n''+1)}
         {\Gamma(a+b)} \: .
\label{fermeture-ter}
\end{equation}
If one normalizes (\ref{fermeture}) by (\ref{norm}) one gets
 \begin{align}
  \frac{\langle \rho_a,a|Y_1 Y_2|\rho_b,b\rangle}
        {\langle \rho_a,a| (D+E)^{L+L'}    |  \rho_b,b\rangle}
   = {\Gamma(L+L'+a+b) \over \Gamma(L+a+b) \; \Gamma(L'+1)}
   \oint\displaylimits_{\rho_b<|\rho|<\rho_a} &  \frac{d\rho}{2i\pi}\:
     \frac{(\rho_a-\rho_b)^{a+b+L+L'}}{(\rho_a-\rho)^{a+b+L}(\rho-\rho_b)^{1+L'}}
 \ \times
 \label{fermeture-bis1}
 \\ \nonumber 
 &     \frac{\langle \rho_a,a|Y_1|\rho, b\rangle}
          {\langle \rho_a,a| (D+E)^L |  \rho, b\rangle} \:
     \frac{\langle \rho,1-b|Y_2|\rho_b,b\rangle}
          {\langle \rho,1-b|  (D+E)^{L'} | \rho_b,b\rangle} \  .
 \end{align}

An additivity relation more general than (\ref{fermeture})  can be proved
for the ASEP  
 \cite{DLS4}. The special case of the TASEP will be discussed in section 
\ref{add-tasep}  below. 

\section{Large deviation function of density profiles}
\label{ld-dens-ssep}
If one divides  a chain of  $L$ sites  into  $n$ boxes of linear
size $l$ (there are  of course $n = L/l$ such boxes), one can
try to determine the probability of finding a certain density profile $\{
\rho_1, \rho_2, ... \rho_n\}$, i.e. the probability of seeing $l
\rho_1$ particles in the first box, $l \rho_2$
particles in the second box, ... $l \rho_n$ in the $n$th box. For large
$L$ one expects (\ref{finite}) the following $L$ dependence of this probability
\begin{equation}
{\rm Pro}_L(\rho_1,... \rho_n|\rho_a,\rho_b) \sim \exp[ - L {\cal F}_n(\rho_1, \rho_2,
...\rho_n|\rho_a,\rho_b)] \ .
\label{finite-bis}
\end{equation}
If one defines a reduced coordinate $x$ by
\begin{equation}
i= L x
\end{equation}
and if one takes the limit $l \to \infty$ with $l \ll L$ so that the
number of boxes becomes infinite,  one gets as in (\ref{continu})  the
large deviation functional ${\cal
F}(\rho(x))$
\begin{equation}
{\rm Pro}_L(\{\rho(x)\}) \sim \exp[ - L {\cal F}(\{\rho(x)\}|\rho_a,\rho_b)]  \ .
\label{LDdef}
\end{equation}
For the SSEP (in one dimension), the  functional   ${\cal
F}(\rho(x)|\rho_a,\rho_b)$  is
given by the following exact expressions:
\\ \\
{\bf at equilibrium,} i.e. for $\rho_a=\rho_b=r$
\begin{equation}
{\cal F}(\{\rho(x)\}|r,r)=\int_0^1 B(\rho(x),r) dx
\label{F1}
\end{equation}
 where
\begin{equation}
B(\rho,r) = (1- \rho) \log {1- \rho \over 1 - r} \  +\  \rho \log {\rho
\over r}
\; .
\label{Br}
\end{equation}
This can be   derived easily.  When $\rho_a=\rho_b=r$, the steady state
is a Bernoulli measure (\ref{equilibrium}) where
all the sites are occupied independently  with probability $r$.
Therefore if one divides a chain of length $L$ into $L/l$
intervals of length $l$,  one has
\begin{equation}
{\rm Pro}_L(\rho_1,... \rho_n|r,r)=  \prod_i^{L/l} {l! \over [l \rho_i]! \ \
[l(1-\rho_i)]!} \   r^{l\rho_i} \  (1-r)^{l(1-\rho_i)}
\end{equation}
and using Stirling's formula one gets  (\ref{F1},\ref{Br}).
\\ \\
{\bf For the non-equilibrium case,} i.e. for $\rho_a \neq\rho_b$, it was
shown in \cite{DLS1,BDGJL2,DLS2} that
\begin{equation}
{\cal F}(\{\rho(x)\}|\rho_a,\rho_b)= \int_0^1 dx  \ \left[ B(\rho(x),F(x)) + \log{F'(x)
\over \rho_b - \rho_a}  \right]
\label{F2}
\end{equation}
where the function $F(x)$ is the monotone solution of the differential
equation
\begin{equation}
\rho(x) = F + {F(1-F) F'' \over F'^2 }
\label{F3}
\end{equation}
satisfying the boundary conditions $F(0)= \rho_a$ and $F(1)= \rho_b$.
This expression shows that ${\cal F}$ is a {\it non-local} functional  of
the density  profile $\rho(x)$ as
 $F(x)$ depends  on the profile $\rho(y)$ at all
points $y$.
For example if the difference $\rho_a-\rho_b$ is small, one can expand
${\cal F}$
and obtain  the expression (\ref{secondorder}) where the non-local character of the functional
is clearly visible: at second order in $\rho_a-\rho_b$, 
 one gets by solving  (\ref{F3})
\begin{align}
F  = &   \rho_a - (\rho_a- \rho_b) x - {(\rho_a - \rho_b)^2 \over \rho_a (1-\rho_a)}
\left[ (1-x) \int_0^x y (\rho(y)- \rho_a) dy + x \int_x^1 (1-y) (\rho(y)-
\rho_a) dy  \right] + O \left((\rho_a-\rho_b)^3\right)
\nonumber \\
 = &  \rho^*(x)  - {(\rho_a - \rho_b)^2 \over \rho_a (1-\rho_a)}
\left[ (1-x) \int_0^x y (\rho(y)- \rho^*(x)) dy + x \int_x^1 (1-y)
(\rho(y)- \rho^*(x)) dy  \right]
+ O \left((\rho_a-\rho_b)^3\right)
\label{rhoa-rhob-small}
\end{align}
and this leads to (\ref{secondorder}) by replacing into (\ref{F2}). 
\\ \ \\
{\bf Derivation of (\ref{F2},\ref{F3}):}
in the original derivation of
(\ref{F2},\ref{F3}) from the matrix ansatz \cite{DLS1,DLS2}
the idea was to decompose the chain into $L/l$ boxes of $l$ sites and to
sum the weights given by the matrix ansatz (\ref{matrix},\ref{algebra}) over all the microscopic
configurations for which the number of particles is $l \rho_1$ in the
first box, $l \rho_2$  in the second box ..., $l \rho_n$  in the $n$th box.

An  easier way of deriving (\ref{F2},\ref{F3}) is to write (we do it
here in the particular case where $a+b=1$, i.e. ${1 \over \alpha + \gamma}
+ {1 \over \beta + \delta}=1$, and $\rho_b
< \rho_a$) from  (\ref{fermeture}) and (\ref{norm}) where $Y_1$ and $Y_2$ represent sums over all configurations with $k l$ sites with a density $\rho_1$ in the first $l$ sites, ... $\rho_k$ in the $k$-th  $l$ sites, and $Y_2$ a similar sum for the $(n-k) l$ remaining sites.

\begin{align}
P_{n l} (& \rho_1, \rho_2 ... \rho_n | \rho_a ,\rho_b) = {(k l)! \;  ((n-k)l)!
\over (nl)! }
\oint\displaylimits_{\rho_b<|\rho|<\rho_a}  {d \rho \over 2 i \pi}
\times
\label{PPP}
\\
\nonumber
&{(\rho_a - \rho_b)^{nl+1} \over (\rho_a - \rho)^{kl+1} (\rho-
\rho_b)^{(n-k)l+1}}
P_{k l} ( \rho_1 ... \rho_k | \rho_a, \rho)
\ P_{(n-k) l} ( \rho_{k+1} ... \rho_n | \rho, \rho_b)
\end{align}
Note that in (\ref{PPP}) the density $\rho$ has  become a complex variable. This is not  a difficulty as all the weighs (and therefore the probabilities which appear in (\ref{PPP})) are rational functions of $\rho_a$ and $\rho_b$.

For large $nl$, if one writes $k=nx$, one gets by evaluating (\ref{PPP}) at the saddle point 
\begin{align}
{\cal F}_n(\rho_1, \rho_2, ...\rho_n|\rho_a,\rho_b)
= & \max_{\rho_b < F < \rho_a}
x{\cal F}_k(\rho_1,  ...\rho_k|\rho_a,F)
+(1-x){\cal F}_{n-k}(\rho_{k+1},  ...\rho_n|F,\rho_b)
\nonumber \\
& +x \log \left({\rho_a -F \over x}\right)
+(1-x) \log \left({F-\rho_b  \over 1-x}\right)
-\log(\rho_a-\rho_b)
\end{align}
(To estimate  (\ref{PPP}) by a saddle point method, one should find the value $F$  of $\rho$ which maximizes  the integrand over the contour. As the contour is perpendicular to the real axis at their crossing point, this becomes a minimum when $\rho$ varies along the real axis).
If one repeats the same procedure $n$ times, one gets
\begin{equation}
\label{Fn}
{\cal F}_n(\rho_1, \rho_2, ...\rho_n|\rho_a,\rho_b)
=
 \max_{\rho_b=F_0 < F_1  ..  <F_i < ..< F_n=\rho_a}
{1 \over n}\sum_{i=1}^n{\cal F}_1(\rho_i|F_{i-1},F_i) + \log \left( {(F_{i-1}-F_i)n \over \rho_a - \rho_b}\right)
\end{equation}
For large $n$, as $F_i$ is monotone, the difference $F_{i-1}-F_
i$ is small for almost all $i$ and one can replace ${\cal F}_1(\rho_i|
F_{i-1},F_i) $ by its equilibrium value ${\cal F}_1(\rho_i|F_{i},F_i) = B(\rho_i,F_i)$. If one write  $F_i$ as a function of $i/n$
\begin{equation}
F_i = F\left({i \over n} \right)
\end{equation}
 (\ref{Fn}) becomes  
\begin{equation}
{\cal F}(\{\rho(x)\}|\rho_a,\rho_b)= \max_{F(x)}\int_0^1 dx  \ \left[ B(\rho(x),F(x)) + \log{F'(x)
\over \rho_b - \rho_a}  \right]
\label{F5}
\end{equation}
where the maximun is over all the monotone  functions $F(x)$ which satisfy
$F(0=\rho_a$ and $F(1)=\rho_b$ and one gets  (\ref{F2},\ref{F3}).
 \\ \ \\
{\bf Remark:}
One can easily get from (\ref{F2},\ref{F3})  the generating function 
${\cal G}(\{\alpha(x)\})$ 
of the density (\ref{Gdef}) for the SSEP:
\begin{equation}
{\cal G}(\{\alpha(x)\})= \int_0^1 dx \left[ \log( 1-F + F e^{\alpha(x)} ) \ - \ \log {F' \over \rho_b - \rho_a } \right]
\label{Ga}
\end{equation}
where $F$ is the monotone solution of
\begin{equation}F'' + { F'^2 (1- e^{\alpha(x)}) \over 1-F + F e^{\alpha(x)}} =0 
\label{Falpha}
\end{equation}
with $F(0)= \rho_a$ and $F(1)=   \rho_b$. For small $\alpha(x)$  the
solution of (\ref{Falpha}) is to second order in  the difference $\rho_a- \rho_b$ 
\begin{equation}
F(x)= \rho^*(x) 
   - (\rho_a - \rho_b)^2 
\left[ (1-x) \int_0^x y \; \alpha(y)\;   dy + x \int_x^1 (1-y) \;
 \alpha(y)  \; dy  \right] \ .
\end{equation}
This leads  to  ${\cal G}(\alpha(x))$ at   order  $(\rho_a-\rho_b)^2$
\begin{equation}
{\cal G}(\{\alpha(x)\}) = \int_0^1 dx \left[\rho^*(x) \alpha(x) + {\rho^*(x)(1-\rho^*(x)) \over 2} \alpha(x)^2 \right] - (\rho_a-\rho_b)^2 \int_0^1 dx \int_x^1 dy  \ x (1-y)  \ \alpha(x) \alpha(y)  
\end{equation}
and
one recovers through (\ref{2pt}) the expression of the two-point correlation function (\ref{2pt-function}).

\section{The macroscopic fluctuation theory}
\label{mft}

For a general diffusive one dimensional system  (figure \ref{reservoir}) of linear size $L$
the average current and the  fluctuations of this current  
 near equilibrium can be characterized by  
two quantities $D(\rho)$ and $\sigma(\rho)$ defined by
\begin{equation}
\lim_{t \to \infty}{\langle Q_t \rangle \over t}=  {D(\rho) \over L} (\rho_a - \rho_b)
\ \ \ \ \ \ \ {\rm for \ } 
\ (\rho_a - \rho_b) \ \   {\rm small}
\label{Ddef}
\end{equation}
\begin{equation}
\lim_{t \to \infty}{\langle Q_t^2 \rangle \over t}  = {\sigma(\rho) \over L}
\ \ \ \ \ \ \ {\rm for \ } 
\ \rho_a = \rho_b 
\label{sigmadef}
\end{equation}
where $Q_t$ is the total number of particles transferred from the left
reservoir to the system during time $t$.

Starting from the hydrodynamic large deviation theory \cite{KOV,Spohn,KL}
Bertini, De Sole, Gabrielli, Jona-Lasinio and Landim
\cite{BDGJL1,BDGJL2,BDGJL3} have developed a   general
approach, {\it the macroscopic fluctuation theory}, to calculate the large deviation functional $\cal F$ of the density (\ref{continu}) in the 
non-equilibrium steady state of a  system in contact with two (or more)
reservoirs as in figure \ref{reservoir}.
Let us briefly sketch their approach.
For diffusive systems (such as the SSEP),
the density $\rho_i(t)$ near position $i$ at time $t$ and
 the total flux $Q_i(t)$ flowing through position $i$ between time $0$ and time $t$  are
for a large system of size $L$
and for times of order $L^2$, scaling functions of the form
\begin{equation}
 \rho_i(t) =  \widehat \rho \left( {i\over L}, {t \over L^2} \right) \; , 
\qquad {\rm and} \qquad
 Q_i(t) = L \widehat Q \left( {i\over L}, {t \over L^2} \right) \; 
\end{equation}
(Note that due to the conservation of the number of particles, the scaling form of $\rho_i(t)$ implies the scaling form of $Q_i(t)$).
If one  introduces the instantaneous (rescaled) current defined  by
\begin{equation}
\widehat j(x,\tau) = {\partial \widehat Q (x,\tau) \over \partial \tau}
\end{equation}
 the conservation of the number of particles implies that
\begin{equation} {
\partial \widehat{\rho}(x, \tau) \over \partial \tau}=
-{\partial^2 \widehat{Q} (x,\tau) \over \partial \tau \partial x }
= -{\partial \widehat{j}(x,\tau) \over   \partial x } \, .
\label{conservation}
\end{equation}
Note that
the total flux of particles
through position $i= [L x]$ during the macroscopic time interval $d \tau$, i.e. during the microscopic time interval
$L^2 d\tau$, is  
 $L \widehat{j}(x,\tau) d \tau$. Thus
the microscopic current  is  of order $1/L$ while   the rescaled current $\widehat j$ remains of order $1$.

The {\it macroscopic fluctuation theory} \cite{BDGJL1,BDGJL2,BDGJL3} starts from
the probability of observing
a certain density profile $\widehat \rho \left( x, \tau \right)$ and 
current profile $\widehat j \left( x, \tau \right)$
over the rescaled time interval $\tau_1 < \tau < \tau_2$
\begin{equation}
\label{eq: dev exp}
Q_{\tau_1,\tau_2} \Big( \{\widehat \rho(x,\tau), \widehat j(x,\tau)\}  \Big)  \sim \exp
\left[ - L \int_{\tau_1}^{\tau_2}
d \tau' \int_0^1 dx {\left[\widehat j(x,\tau') + D({\widehat \rho(x,\tau'}))
{\partial {\widehat \rho(x,\tau')} \over \partial x}\right]^2 \over 2
\sigma(\widehat \rho(x,\tau'))} \right]
\end{equation}
 where the current $\widehat j(x,s)$ is related to the density profile $\widehat \rho(x,s)$
by the conservation law (\ref{conservation}) 
and
the functions $D(\rho)$ and $\sigma(\rho)$ are defined by (\ref{Ddef},\ref{sigmadef}).  Note that a similar expression was obtained in \cite{PJSB,JSP} 
by considering stochastic models in the context of shot noise in
mesoscopic quantum conductors.
 
Then  Bertini et al  \cite{BDGJL1} show   that to calculate the probability of  observing a density profile $\rho(x)$ in the steady state,
 one has to find out how this deviation is produced. For large $L$,
one has to find  the  optimal path $\{\widehat{\rho}(x,s),\widehat{j}(x,s)\}$ for $-\infty < s < \tau$  in
the space of density and current profiles 
 and
\begin{equation}
{\rm Pro}(\rho(x)) \sim \max_{ \{\widehat{\rho}(x,s), \widehat{j}(x,s) \} }
{\rm Q}_{-\infty,\tau}\Big(\{ \widehat{\rho}(x,s), \widehat{j}(x,s) \}\Big)
\end{equation}
which goes from the typical profile $\rho^*(x)$ to the desired profile
\begin{equation}
 \widehat{\rho}(x,-\infty) = \rho^*(x)  \ \ \ \ \ ; \ \ \ \
 \widehat{\rho}(x,\tau) = \rho(x) \ . \label{bc}
\end{equation}
This means that the large deviation functional ${\cal F}$ of the density (\ref{LDdef}) is given by
\begin{equation}
{\cal F}(\rho(x))= \min_{\{ \widehat{\rho}(x,s), \widehat{j}(x,s) \}}
\int_{-\infty}^{\tau}
d \tau' \int_0^1 dx { \left[\widehat{j}(x,\tau') + D({\widehat{\rho}(x,\tau'}))
{\partial {\widehat{\rho}(x,\tau')} \over \partial x}\right]^2 \over 2
\sigma(\widehat{\rho}(x,\tau'))} 
\label{FBertini}
\end{equation}
where the density  and the current profiles satisfy  the conservation law (\ref{conservation})  and the  boundary conditions (\ref{bc}).

Finding this optimal path   $\widehat{\rho}(x,s),  \widehat{j}(x,s)$ with
the boundary conditions (\ref{bc})  is usually a hard problem. 
Bertini et al  \cite{BDGJL1} were however able to write an  equation satisfied by
${\cal F}$: as (\ref{FBertini}) does not depend on $\tau$, one can rewrite  it as
\begin{equation}
{\cal F}(\rho(x))= \min_{ \delta \rho(x), j(x)} \left[ 
{\cal F}(\rho(x)-\delta \rho(x))  
+ \delta  \tau \int_0^1 dx { \left[j(x) + D(\rho(x))
\rho'(x) \right]^2 \over 2
\sigma(\rho(x))}
\right]
\label{FBertini-bis}
\end{equation}
where $\rho(x)-\delta \rho(x)=\widehat{\rho}(x,\tau-d\tau)$
and $j(x) = \widehat{j}(x,\tau)$. 
Then if one defines $U(x)$ by
\begin{equation}
U(x) = {\delta {\cal F}(\{\rho(x)\}) \over \delta \rho(x)}
\label{Udef}
\end{equation}
and one uses the conservation law
$\delta \rho(x)= -{dj(x) \over dx} d \tau$
one should have according to (\ref{FBertini-bis})
that the optimal current $j(x)$ is given by
\begin{equation}
j(x)= -D(\rho(x)) \rho'(x) + \sigma(\rho(x)) U'(x) \ .
\label{j-bertini}
\end{equation}
 Therefore starting with $\widehat{\rho}(x,\tau)=\rho(x)$ and using the
time evolution 
\begin{equation}
 {d \widehat{\rho}(x,s) \over ds} = - {d \widehat{j}(x,s) \over dx} 
\label{dyn}
\end{equation}
with $\widehat{j}$ related to $\widehat{\rho}$ by (\ref{j-bertini}) one
should get  the whole time dependent optimal profile
$\widehat{\rho}(x,s)$ which converges to $\rho^*(x)$ in the limit $s \to -\infty$. The problem of course is that $U(x)$  defined in (\ref{Udef}) is not known.
One way of thinking of the problem is to say that for every $\rho(x)$, the function $U(x)$ should be adjusted so that the dynamics (\ref{dyn},\ref{j-bertini}) gives $\widehat{\rho}(x,s) \to \rho^*(x)$ as $s \to - \infty$.

One can write from 
 (\ref{FBertini-bis})   (after an integration by part and using the fact that $U(0)=U(1)=0$ if $\rho(0)=\rho_a$ and $\rho(1)=\rho_b$) the equation satisfied by $U'(x)$ 
\begin{equation}
 \int_0^1 dx \left[ \left({D \rho' \over \sigma} - U' \right)^2 - \left({D \rho' \over \sigma  }\right)^2  \right] {\sigma \over 2}=0
\label{hamilton-jacobi}
\end{equation}
which is the Hamilton-Jacobi \cite{BDGJL1} equation of Bertini et al. 
For general $D(\rho)$ and $\sigma(\rho)$ one does not know how to find
the solution $U'(x)$ of (\ref{hamilton-jacobi})
 for an arbitrary $\rho(x)$ and thus
 one does not know how to get   a more explicit expression of the large deviation function ${\cal F}(\{\rho(x)\})$.

One can however  check rather easily whether a given expression of ${\cal F}(\{\rho(x)\})$
satisfies (\ref{hamilton-jacobi}) since $U'(x)$ can be calculated from 
(\ref{Udef}). For the SSEP one gets
from (\ref{F5},\ref{Udef})
\begin{equation}
U(x) = \log \left[ {\rho(x) (1- F(x)) \over (1-\rho(x))F(x)} \right]
\end{equation}
with $F(x)$ related to $\rho(x)$  by (\ref{F3}).
One can then check that (\ref{hamilton-jacobi}) is indeed satisfied using
the expressions of $D=1$ and $\sigma= 2 \rho(1-\rho)$ for the SSEP (see
(\ref{sigmaSSEP}) below).

In fact  
${\cal F}$ is known, one can obtain the whole optimal path
$\widehat{\rho}(x,s)$ from the evolution (\ref{dyn}) with $\widehat{j}$  related to $\widehat{\rho}$ by (\ref{j-bertini}) which becomes for the SSEP
\begin{equation}
\label{j-bertini1}
\widehat{j}(x,s)= - {d \widehat{\rho}(x,s) \over dx} + 
\sigma (\widehat{\rho}(x,s))   \log \left[ {\widehat{\rho}(x,s)  (1- \widehat{F}(x,s))  \over (1-\widehat{\rho}(x,s))  \widehat{F}(x,s) } \right]
\end{equation}
where $\widehat{F}$ is related to $\widehat{\rho}$ by (\ref{F3}).
For  (\ref{F2},\ref{F3}) to coincide with (\ref{FBertini}), the
optimal profile $\widehat{\rho}$ evolving according to (\ref{dyn}) 
should  converge to $\rho^*(x)$ as $s \to -\infty$.
One can check  that this evolution of $\widehat{\rho}(x,s)$ is equivalent
to the following evolution (\cite{BDGJL2}) of $\widehat{F}$ 
\begin{equation}
\label{F-bertini}
{d \widehat{F}(x,s) \over ds} = - {d^2 \widehat{F}(x,s) \over ds}
\end{equation}
where $\widehat{F}$ is related to $\widehat{\rho}$ by (\ref{F2}).
Clearly $\widehat{F}(x,s) \to \rho^*(x)$ and therefore $\widehat{\rho}(x,s)
\to \rho^*(x)$ as $s \to - \infty$. Thus (\ref{dyn},\ref{j-bertini1})
 do give the optimal path  in (\ref{FBertini}) with
the right boundary conditions (\ref{bc})
and   (\ref{FBertini}) coincides for the SSEP with the prediction
(\ref{F2},\ref{F3}) of the matrix approach.

Apart for the SSEP, the large deviation functional $\cal F$ of the
density is so far known only in very few cases: the Kipnis Marchioro
Presutti  model \cite{KMP,BGL}  the  weakly asymmetric exlcusion process
\cite{ED,DPS} and the ABC model \cite{CDE,EKKM} on a ring.

\section{Large deviation of the current and the fluctuation theorem}
\label{ldc}
For a system in contact with two reservoirs at densities $\rho_a$ and $\rho_b$, 
as in figure \ref{reservoir},
one can try to study the probability distribution \cite{HRS1} of the  total number $Q_t$ of particles  which flows through the system during time $t$.
For finite $t$, this  distribution depends on the initial condition of
the system as well as on the place where the flux $Q_t$ is measured (along an arbitrary
section of the system,
at the boundary with the left reservoir or at the boundary with the right reservoir). In the long time limit however, if the system has
a finite relaxation time and if the number of particles in the system is
bounded (i.e. infinitely many particles cannot accumulate in the system) 
the probability distribution of $Q_t$ takes the form 
\begin{equation}
{\rm Pro}\left( {Q_t \over t} = j \right) \sim e^{-t F(j)}
\label{F(j)}
\end{equation}
where 
 the large deviation function $F(j)$ of the current $j$ depends neither
on the initial condition nor on where the flux $Q_t$ is measured. This
large deviation function $F(j)$ has  usually a shape similar to
$a_{\vec{r}}(\rho)$ in figure \ref{ar}, with a minimum the typical  value $j^* = \langle J \rangle $ (the avergae current) where $F(j^*)=0$.

It is often as convenient to work with  the generating function $\left\langle e^{\lambda Q_t} \right\rangle$. In the long time limit
\begin{equation}
\left\langle e^{\lambda Q_t} \right\rangle \sim e^{ \mu(\lambda) \; t }
\label{mu(lambda)}
\end{equation}
where $\mu(\lambda)$ is  clearly the Legendre transform of the large deviation function $F(j)$
\begin{equation}
\label{legendre1}
\mu(\lambda) = \max_j[ \lambda j - F(j)]
\end{equation}
As in section \ref{non-loc}  , the knowledge of $\mu(\lambda)$ determines the cumulants of $Q_t$
\begin{equation}
\lim_{t \to \infty} {\langle Q_t^k\rangle _c \over t} =  \left. {d^k \mu(\lambda) \over d \lambda^k} \right|_{\lambda=0}
\label{cum}
\end{equation}
when the expansion in powers of $\lambda$ is justified.
\\ \ \\
According to the {\it fluctuation theorem,}
\cite{ECM,GC,ES,Kurchan,LS,MRW,Seifert,G,Crooks,M1,G2,Gaspard}
 the large deviation function $F(j)$ of the current  satisfies the following symmetry property
\begin{equation}
\label{fluctuation-theorem-3}
 F(j) -  F(-j) = - j   [ \log z_a - \log  z_b ]
\qquad {\rm and} \qquad
\mu(\lambda)=  \mu\left( - \lambda +  \log z_b  - \log z_a \right) \  .
\end{equation}
{\bf Proof:} Following previous derivations  
\cite{Kurchan,LS,M1}
for  stochastic dynamics 
 the fluctuation theorem 
 (\ref{fluctuation-theorem-3})
 can be easily recovered  \cite{BD4} from
the {\it generalized detailed balance relation} (\ref{generalized-detailed-balance-bis}).
This can be seen by  comparing the probabilities of a
trajectory in phase space and of its time reversal for a system in contact with two reservoirs.
A trajectory $"Traj"$ is specified by a sequence of successive
configurations $C_1,... C_k$ visited by the system,
the times $t_1,...t_k$ spent in each of these configurations, and the number of particles  $
q_{a,i}, q_{b,i}$ transferred from the reservoirs  to the system  when the
system jumps from $C_i$ to $C_{i+1}$.
\[ {\rm Pro}(Traj) =  dt^{k-1}  \
\left[\prod_{i=1}^{k-1}  W_{q_{a,i}, q_{b,i}} (C_{i+1},C_i) \right]
 \ \exp \left[- \sum_{i=1}^{k} t_i  \ r(C_i)    \right] \]
where $r(C) = \sum_{C'} \sum_{q_a,q_b} W_{q_{a}, q_{b}} (C',C) $ and $dt$ is the infinitesimal
time interval over which jumps occur.

For the trajectory $"-Traj"$ obtained  from $"Traj"$ by time reversal,
i.e. for which the system visits successively the configurations
$C_k,... C_1$,  exchanging  $-q_{a,i}, -q_{b,i}$ particles with the reservoirs  each time
the system jumps from $C_{i+1}$ to $C_i$, one has
\[ {\rm Pro}(-Traj) =  dt^{k-1}  \
\left[\prod_{i=1}^{k-1}  W_{-q_{a,i}, -q_{b,i}} (C_{i},C_{i+1}) \right]
 \ \exp \left[- \sum_{i=1}^{k} t_i \ r(C_i)    \right] \]
One can see from the generalized detailed balance relation
(\ref{generalized-detailed-balance-bis}) that 
\begin{equation}
 {{\rm Pro}(Traj) \over {\rm  Pro}(-Traj) } = \exp \left[  \sum_{i=1}^{k-1}
q_{a,i}  \log z_a -  q_{b,i} \log z_b \right] = \exp \left[ 
Q_t^{(a)} \log z_a -  Q_t^{(b)}  \log z_b \right]
\end{equation}
where $Q_t^{(a)}= \sum_i q_{a,i} $ and $Q_t^{(b)}= \sum_i q_{b,i} $
are the total number of particles transferred from the reservoirs $a$ and $b$ to the
system during time $t$.

In general $Q_t^{(a)} $ and $Q_t^{(b)} $ grow with time but their sum remains bounded (if one assumes that  particles cannot accumulate in the system). Therefore for large time $Q_t \equiv Q_t^{(a)}= -  Q_t^{(b)} +o(t)$ 
and
\begin{equation}
 {{\rm Pro}(Traj) \over {\rm  Pro}(-Traj) } \sim  
 \exp \left[ 
Q_t (\log z_a  - \log z_b )\right]
\label{ratio}
\end{equation}
Summing over all trajectories \cite{BD4}, taking the log and then the long time
limit (\ref{F(j)}) leads   to the  fluctuation
theorem (\ref{fluctuation-theorem-3}).
\\ \ \\
{\bf Remark:}
The fluctuation theorem predicts  a symmetry relation similar to
(\ref{fluctuation-theorem-3}) for the heat current  for a system in
contact with two heat baths at unequal temperatures as in figure
\ref{heatbath}. Under similar conditions as for the current of particles
(the energy of the system is bounded and the relaxation time is finite -
 see
 \cite{farago,HRS2,visco} for counter-examples where the energy is not bounded in
which case the fluctuation theorem has to be modified)
one gets that the distribution of the energy $Q_t$ flowing through the
system during a long time $t$ is given by (\ref{F(j)}) and that the large
deviation function $F(j)$ or its Legendre transform satisfy
\begin{equation}
F(j) -  F(-j) = -j \left( {1 \over kT_b} - {1 \over kT_a}\right)
\ \ \ \ \ ; \ \ \ \ \
\mu(\lambda)=  \mu\left( - \lambda + {1 \over k  T_a} - {1 \over  k T_b}\right)
\label{fluctuation-theorem-1}
\end{equation}
which states that the difference ${\cal} F(j) - {\cal} F(-j)$ is linear in $j$ with a universal slope
related to the difference of the inverse temperatures. Note that $j ( {1
\over kT_b} - {1 \over kT_a}) $ is the rate of entropy production which
is the  quantity generally used to state the fluctuation theorem
\cite{ECM,GC,LS,Seifert}.

\section{The fluctuation-dissipation theorem}
\label{fdt}
In the limit of small $T_a- T_b$ (i.e. close to equilibrium), one can recover from
(\ref{fluctuation-theorem-1}) the {\it fluctuation-dissipation relation}
between 
the response to a small temperature gradient
\begin{equation}
 {\langle Q_t \rangle \over t } \to (T_a - T_b) \widetilde{D} \ \ \ \ \ \
{\rm for} \  \ T_a-T_b  \ \ {\rm small}
\label{D-def}
\end{equation}
and
the variance of the energy flux at equilibrium
\begin{equation}
 {\langle Q_t^2 \rangle \over t } \to \widetilde{\sigma} \ \ \ \ \ \ {\rm
for} \  T_a=T_b \  .
\label{sigma-def}
\end{equation}
In fact from these definitions of $\widetilde{D}$ and $\widetilde{\sigma}$, one has
\begin{equation}
 \mu(\lambda) =   (T_a- T_b)\widetilde{D} \lambda  + {\widetilde{\sigma} \over 2 } \lambda^2 + O \left( \lambda^3,  \lambda^2 (T_a - T_b),  \lambda (T_a - T_b)^2 \right)
\label{fluctuation-dissipation}
\end{equation}
and   using the fluctuation theorem (\ref{fluctuation-theorem-1}), one
gets that the coefficients $\widetilde{\sigma}$ and $\widetilde{D}$ have to satisfy
\begin{equation}
\widetilde{\sigma} = 2 k T_a^2 \widetilde{D}
\label{fluctuation-dissipation-bis}
\end{equation}
 which is the usual  fluctuation-dissipation relation.
In general both $\widetilde{D}$ and $\widetilde{\sigma}$ depend on the temperature $T_a$.
\\ \ \\
 The same close-to-equilibrium expansion of (\ref{fluctuation-theorem-3}) for  a
current of particles leads to 
\begin{equation}
  \widetilde{\sigma} = 2 
 {d \rho \over d \log z}
\widetilde{D}
= 2 k T \rho^2 \kappa(\rho)
\widetilde{D}
\label{fluctuation-dissipation-ter}
\end{equation}
where the coefficients $\widetilde{\sigma}$ and $\widetilde{D}$ are defined  as in (\ref{sigma-def},\ref{D-def}) by
\begin{equation}
 {\langle Q_t \rangle \over t } \to (\rho_a - \rho_b) \widetilde{D}  \ \ \ \ {\rm for} \  \rho_
a-\rho_b  \ {\rm small}
\ \ \ \ \ \ \ \ \ \ \ \ {\rm and} \ \ \ \ \ \ \
{\langle Q_t^2 \rangle \over t } \to \widetilde{\sigma} \ \ \ \ \ \ {\rm for} \  \rho_a=\rho_b
\label{sigma-def-bis}
\end{equation}
and where $\kappa(\rho)$ 
is the compressibility  (\ref{compressibility}) at equilibrium.

To see why the compressibility appears in
(\ref{fluctuation-dissipation-ter}),  one can write $\log Z  =-F/k T = -V
f(N/V)/k T$  where $F$ is the total free energy and $f$ the free energy
per unit volume. One then uses the facts that the fugacity $z$ is given by  $k T \log z   = d F/dN= f'(\rho) $, that
$p= - dF/dV= \rho f'(\rho) - f(\rho)$ and  thus $d \rho /d \log z= k T/f''(\rho)=k T \rho d\rho / dp = kT \rho^2 \kappa(\rho)$.
 \\ \ \\
In the case of the SSEP, one has from (\ref{current},\ref{Ddef}) that  $D_{\rm SSEP}=1$. As the free energy $f(\rho)$ of the SSEP (at equilibrium at density $\rho$) is 
\begin{equation}
f(\rho)= kT [\rho \log \rho + (1-\rho) \log(1- \rho)]
\end{equation}
one has 
\begin{equation}
\label{zdef}
\log z = \log{\rho \over 1- \rho} \ .
\end{equation}
Thus     $d \rho /d \log z= k T/f''(\rho)= \rho(1-\rho)$ and 
thanks to (\ref{fluctuation-dissipation-ter}) and (\ref{current},\ref{Ddef})
one  gets
\begin{equation}
D_{\rm SSEP} = 1 \ \ \ \ \  \ \ ; \ \ \  \ \ \ \
\sigma_{\rm SSEP} = 2 \rho (1- \rho)
\label{sigmaSSEP}
\end{equation}
 \ \\
Note that in (\ref{Ddef},\ref{sigmadef}) there is, 
 compared with (\ref{sigma-def},\ref{D-def},\ref{sigma-def-bis}),
an extra $1/L$ factor in the definition of $\sigma$ and $D$ 
 to get  a finite large $L$ limit of $\sigma$ and $D$. Of course, with this   extra $1/L$ fator, both (\ref{fluctuation-dissipation-bis})
and (\ref{fluctuation-dissipation-ter})
remain valid.

\section{Current fluctuations in the SSEP}
\label{cur-ssep}
For the SSEP, if $Q_t$ is the total number of particles transferred from the left reservoir to the system during  a long time $t$, one has (\ref{mu(lambda)})
\begin{equation}
\left\langle e^{\lambda Q_t} \right\rangle \sim e^{\mu(\lambda)  \; t}
\; .
\end{equation}
The fluctuation theorem (\ref{fluctuation-theorem-3}) implies (\ref{zdef})
a symmetry relation satisfied by $\mu(\lambda)$ 
\begin{equation}
\mu(\lambda) = \mu\left(  - \lambda -\log {\rho_a \over 1- \rho_a} 
+\log {\rho_b \over 1- \rho_b} 
\right) \  .
\label{gc}
\end{equation}
but of course this symmetry  does not determine $\mu(\lambda)$.
 We are now going to see that,
because the evolution is Markovian, 
 $\mu(\lambda)$ can
be determined as the largest eigenvalue of a certain matrix \cite{DDR,DL,DA,GKP}. 

The probability  $P_t(
 C)$  of finding the system in a configuration $ C$ at time $t$ 
evolves according to (\ref{markov})
\begin{equation}
{dP_t( C) \over dt} =\sum_{ C'} 
W( C, C') P_t( C') 
-W( C',C) P_t( C) 
\; .
\end{equation}
Among all the matrix elements $ W( C, C')$, some correspond to exchanges
 of particles with the left reservoir and others represent internal moves in the bulk or exchanges with the right reservoir.
Thus one can decompose the matrix $W( C, C')$ into three matrices
\begin{equation}
W( C, C')= W_1( C, C')+ W_0( C, C')+ W_{-1}( C, C')
\end{equation}
where here the index is the number of particles transferred from the left reservoir to the system during time $dt$, when the system jumps from the configuration ${ C'}$ to the configuration ${ C}$.
One can then show \cite{DL,DA,DDR} that $\mu(\lambda)$ is simply the largest eigenvalue (more precisely the eigenvalue with largest real part) of the matrix $M_\lambda$
defined by
\begin{equation}  M_\lambda( C, C') = e^{\lambda} W_1( C, C')+ W_0( C, C')+ e^{-\lambda} W_{-1}( C, C') - \delta( C, C') \sum_{ C''} W( C'',C) \ . \end{equation}
In fact the joint probability  $P_t( C,Q_t)$ of $ C $ and $Q_t$
evolves acording to
\begin{equation}
{dP_t( C,Q_t) \over dt} =\sum_{ C'}  \sum_{q=-1,0,1}
W_q( C, C') P_t( C',Q_t-q) 
- \sum_{C'} W( C', C) P_t( C,Q_t) 
\; .
\end{equation}
Then if $\widetilde{P}_t( C) = \sum_{Q_t} e^{\lambda Q_t} P_t( C,Q_t) $
one has 
\begin{equation}
{d\widetilde{P}_t( C,Q_t) \over dt} =\sum_{ C'}  M_\lambda(C,C')
\widetilde{P}_t( C') 
\; .
\end{equation}
and this shows that $\mu(\lambda)$ is the eigenvalue with largest real part of the matrix $M_\lambda$.
\\ \ \\
The size of the matrix $M_\lambda$ grows like $2^L$ (which is the total  number of
possible configurations of a chain of $L$ sites). In \cite{DDR}  a
pertubative approach was developed to calculate  $\mu(\lambda$) in powers of $\lambda$. Let us sketch briefly this approach: one can write down exact expressions for the time evolution
 $ \left\langle e^{\lambda Q_t} \right\rangle$  or of 
 $ \left\langle e^{\lambda Q_t}  H(C) \right\rangle$   where $H(C)$ is an arbitrary function of the configuration $C$ at time $t$. For example 
\begin{equation}
Q_{t+dt} = 
\begin{cases}
Q_t \ \ \ \ {\rm with  \ probability}  \ \  1 - \alpha(1-\tau_1)dt - \gamma \tau_1 dt \cr 
Q_t +1 \ \ \ \ {\rm with  \ probability}  \ \   \alpha(1-\tau_1)dt  \cr 
Q_t -1 \ \ \ \ {\rm with \  probability}  \ \   \gamma \tau_1dt  \cr 
\end{cases}
\end{equation}
and therefore
\begin{equation}
 {d  \left\langle e^{\lambda Q_t} \right\rangle \over dt}  =
\alpha (e^\lambda -1) 
   \left\langle (1- \tau_1) e^{\lambda Q_t} \right\rangle 
+ \gamma (e^{-\lambda} -1) 
   \left\langle  \tau_1 e^{\lambda Q_t} \right\rangle  \  .
\label{E1}
\end{equation}
Similarly one can show 
that for $1<i<L$
\begin{equation}
 {d  \left\langle \tau_i e^{\lambda Q_t} \right\rangle \over dt}  =
\alpha (e^\lambda -1) 
   \left\langle (1- \tau_1) \tau_i e^{\lambda Q_t} \right\rangle
+ \gamma (e^{-\lambda} -1) 
   \left\langle  \tau_1 \tau_i e^{\lambda Q_t} \right\rangle   
 +  \left\langle  (\tau_{i+1  } - 2 \tau_i + \tau_{i-1}) e^{\lambda Q_t} \right\rangle 
\label{E2}
\end{equation}
 the cases $i=1$ are $i=L$ being  slightly different
\begin{equation}
 {d  \left\langle \tau_1 e^{\lambda Q_t} \right\rangle \over dt}  =
\alpha e^\lambda  
   \left\langle (1- \tau_1)  e^{\lambda Q_t} \right\rangle
-  \gamma  
   \left\langle  \tau_1  e^{\lambda Q_t} \right\rangle   
 +  \left\langle  (\tau_{2  } -  \tau_1)  e^{\lambda Q_t} \right\rangle 
\label{E3}
\end{equation}
\begin{equation}
 {d  \left\langle \tau_L e^{\lambda Q_t} \right\rangle \over dt}  =
\alpha (e^\lambda -1) 
   \left\langle (1- \tau_1)\tau_L  e^{\lambda Q_t} \right\rangle
-  \gamma (e^{-\lambda}-1) 
   \left\langle  \tau_1 \tau_L e^{\lambda Q_t} \right\rangle   
 +  \left\langle  (\tau_{L-1  } -  \tau_L)  e^{\lambda Q_t} \right\rangle 
 +  \delta \left\langle  (1 -  \tau_L)  e^{\lambda Q_t} \right\rangle 
 -  \beta \left\langle    \tau_L  e^{\lambda Q_t} \right\rangle  \  .
\label{E4}
\end{equation}
In the long time limit $\left\langle  e^{\lambda Q_t} \right\rangle \sim e^{\mu(\lambda) t}$ and one can define a measure $\langle . \rangle_\lambda$ on the configurations $C$
 \begin{equation}
\langle H(C) \rangle_\lambda= \lim_{t \to \infty}{ \left\langle H(C) e^{\lambda Q_t} \right\rangle \over 
\left\langle  e^{\lambda Q_t} \right\rangle}
\label{H(C)}
\end{equation}
From (\ref{E1}-\ref{E4}) one gets
\begin{equation}
 \mu(\lambda)    =
\alpha (e^\lambda -1) 
   \left\langle (1- \tau_1) \right\rangle_\lambda 
+ \gamma (e^{-\lambda} -1) 
   \left\langle  \tau_1  \right\rangle_\lambda 
\label{G1}
\end{equation}
\begin{equation}
 \mu(\lambda) \langle \tau_i  \rangle_\lambda  =
\alpha (e^\lambda -1) 
   \left\langle (1- \tau_1) \tau_i  \right\rangle_\lambda
+ \gamma (e^{-\lambda} -1) 
   \left\langle  \tau_1 \tau_i  \right\rangle_\lambda   
 +  \left\langle  (\tau_{i+1  } - 2 \tau_i + \tau_{i-1})  \right\rangle_\lambda 
\label{G2}
\end{equation}
\begin{equation}
\mu(\lambda)  \langle \tau_1  \rangle_\lambda   =
\alpha e^\lambda  
   \langle (1- \tau_1)   \rangle_\lambda
-  \gamma  
   \left\langle  \tau_1   \right\rangle_\lambda   
 +  \left\langle  (\tau_{2  } -  \tau_1)   \right\rangle_\lambda 
\label{G3}
\end{equation}
\begin{equation}
\mu(\lambda)  \langle \tau_L \rangle_\lambda   =
\alpha (e^\lambda -1) 
   \left\langle (1- \tau_1)\tau_L   \right\rangle_\lambda
-  \gamma (e^{-\lambda}-1) 
   \left\langle  \tau_1 \tau_L  \right\rangle_\lambda   
 +  \left\langle  (\tau_{L-1  } -  \tau_L)   \right\rangle_\lambda 
 +  \delta \left\langle  (1 -  \tau_L)   \right\rangle_\lambda 
 -  \beta \left\langle    \tau_L   \right\rangle_\lambda  \ .
\label{G4}
\end{equation}

 We see that to get $\mu(\lambda)$ at order $\lambda^k$, one needs to
know (\ref{G1}) the one-point function $\langle  \tau_i \rangle_\lambda$  at order $\lambda^{k-1}$, the two point functions $\langle  \tau_i  \tau_j \rangle_\lambda$ 
at order $\lambda^{k-2}$  (see (\ref{G2}-\ref{G4})) and so on up to the $k-$point functions at order $\lambda^0$.
As the steady state weights $P(C)$ for the SSEP are known exactly (section \ref{matrix})
\cite{DEHP,DLS1,DLS2}, all the correlation  functions are known at order $\lambda^0$ and one can truncate the hierarchy at the level of the $k-$point functions.

In \cite{DDR} this perturbation theory 
based on the hierarchy (\ref{G1}-\ref{G4}) 
was developed to calculate
$\mu(\lambda)$ in powers of $\lambda$. 
The main outcome of this
perturbation theory \cite{DDR}  is that $\mu(\lambda)$, which in principle depends
on $L$, $\lambda$ and on the four parameters $\alpha, \beta, \gamma,
\delta$,
takes  for large $L$ a simple form
\begin{equation}
\mu(\lambda) = {1 \over L } R(\omega) + O \left( {1 \over L^2} \right)
\end{equation}
where  $\omega$ is defined by
\begin{equation}
\omega =  (e^\lambda -1) \rho_a + (e^{-\lambda}-1) \rho_b - (e^\lambda
-1) ( 1- e^{-\lambda}) \rho_a \rho_b  \; .
\label{omegadef}
\end{equation}
where $\rho_a$ and $\rho_b$ are given in (\ref{rhoarhobdef}).
The perturbation theory gives up to fourth order in $\omega$
\begin{equation}
R(\omega) = \omega - {\omega^2 \over 3 } + {8 \omega^3 \over 45} - {4 \omega^4 \over 35 } + O(\omega^5) \; .
\label{romega}
\end{equation}

The fact that $\mu(\lambda)$ depends only on $\rho_a$, $\rho_b$ and $\lambda$ through the single
parameter $\omega$  is the outcome of the calculation, but so far there
is no physical explanation  why it is so. However  $\omega$ remains
unchanged under a number of symmetries \cite{DDR} (left-right,
particle-hole, the Gallavotti-Cohen  symmetry (\ref{gc})) implying that $\mu(\lambda)$ remains unchanged as it should under these symmetries.

From the knowledge of $R(\omega)$ up to fourth order in $\omega$, one can
determine \cite{DDR} the  first four  cumulants (\ref{cum}) of the integrated current $Q_t$ for
arbitrary $\rho_a$ and $\rho_b$:
\begin{itemize}
\item
For $\rho_a=1$ and $\rho_b=0$, one finds

\begin{eqnarray}
 {\langle Q_t \rangle \over t} = {1 \over L} + O \left({1\over L^2 }\right)
\label{q1c1} \\
 {\langle Q_t^2 \rangle_{\rm c} \over t}  =   {1 \over 3 L} + O \left({1\over L^
2 }\right)
\label{q2c1} \\
{\langle Q_t^3 \rangle_{\rm c} \over t}  =   {1 \over 15 L} + O \left({1\over L^
2 }\right)
\label{q3c1} \\
{\langle Q_t^4 \rangle_{\rm c} \over t}  =  {-1 \over 105 L} + O
\left({1\over L^2 }\right) \ .
\label{q4c1}
\end{eqnarray}

These cumulants are the same as the ones known for a
different problem of current flow: the case of non-interacting fermions
through a mesoscopic disordered conductor \cite{LLY,BB}. This can be
understood as a theory \cite{JSP} similar to
the macroscopic fluctuation theory of Bertini et al
\cite{BDGJL1,BDGJL2,BDGJL3,BDGJL4} can be
written for these mesoscopic conductors with the same 
$D(\rho)$ and $\sigma(\rho)$ as for the SSEP (\ref{sigmaSSEP}).

\item
For $\rho_a=\rho_b={1 \over 2}$ which corresponds to an equilibrium case with the same density $1/2$ in the two reservoirs, one finds that all odd cumulants vanish as they should and that
\begin{eqnarray}
 {\langle Q_t^2 \rangle_{\rm c} \over t} =  {1 \over 2 L} + O \left({1\over L^2 }\right)
 \\
{\langle Q_t^4 \rangle_{\rm c} \over t}   =  O \left({1\over L^2 }\right)
\; .
\label{q4d}
\end{eqnarray}
\end{itemize}
Because $\mu(\lambda)$ depends on the parameters $\rho_a,\rho_b$ and
$\lambda$ through the single parameter $\omega$, if one knows
$\mu(\lambda)$ for one single choice of $\rho_a$ and $\rho_b$, then (\ref{omegadef},\ref{romega})  determine $\mu(\lambda)$ for all other choices of $\rho_a,\rho_b$.
In \cite{DDR}, it was conjectured that for the particular case
$\rho_a=\rho_b={1 \over 2}$, not only the fourth cumulant vanishes as in
(\ref{q4d}), but also all the higher cumulants vanish, so that
the distribution of $Q_t$ is Gaussian (to leading order in $1/L$).
This fully determines the function $R(\omega)$ to be
\begin{equation}
R(\omega) = \left[ \log \left (\sqrt{1+ \omega} + \sqrt{\omega} \right)  \right]^2 \; .
\label{Fomega}
\end{equation}
One can then check that,  with this expression of $R(\omega)$,  not only (\ref{q1c1},\ref{q4c1})  but all the higher cumulants
of $Q_t$  in the case $\rho_a=1$ and $\rho_b=0$ coincide with those of fermions through mesoscopic conductors
\cite{LLY,DDR}.

\section{The additivity principle}
\label{add}
In \cite{BD1},
 another conjecture, {\it the additivity principle }
 based on   a simpler physical interpretation,   was formulated which leads for the SSEP to the same expression (\ref{omegadef},\ref{Fomega})
as predicted in section \ref{cur-ssep}   and can be generalized to obtain $F(j)$ or $\mu(\lambda)$ for more general  diffusive systems.

\begin{figure}[ht]
\centerline{\includegraphics[width=10cm]{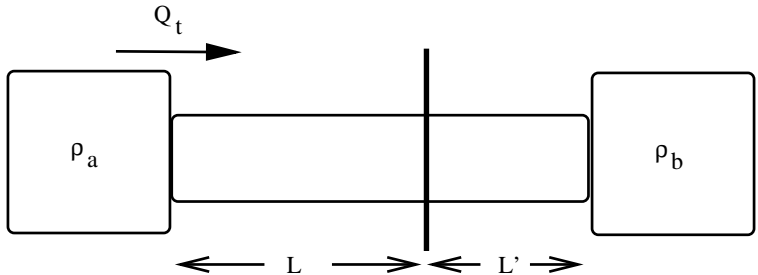}}
\caption{In the additivity principle one tries to relate the large deviation function of the current $F_{L+L'}(j)$ of a system of size $L+L'$ to the large deviation functions $F_L(j)$ and $F_{L'}(j)$ of its two subsystems (\ref{FFF}).}
\label{heatbath1}
\end{figure}
For a system of length $L+L'$  in contact with two reservoirs of
particles at densities $\rho_a$ and $\rho_b$, the probability of
observing, during a long time $t$,  an integrated current $Q_t=jt$ has
the following form (\ref{F(j)})
\begin{equation}
{\rm Pro}_{L+L'}\left(j,\rho_a,\rho_b\right) \sim e^{- t F_{L+L'}\left(j,\rho_a,\rho_b\right)} \; .
\label{FL}
\end{equation}
The idea of the additivity principle is to try to relate the large deviation function $F_{L+L'}(j,\rho_a,\rho_b)$ of the current to the large deviation functions of subsystems by writing that for large $t$
\begin{equation}
{\rm Pro}_{L+L'}\left(j,\rho_a,\rho_b\right) \sim
\max_\rho \left[
{\rm Pro}_{L}\left(j,\rho_a,\rho\right) \times
{\rm Pro}_{L'}\left(j,\rho,\rho_b\right) \right] \; .
\label{PPP1}
\end{equation}
This means that the probability of  transporting a current $j$ over a
distance $L+L'$ between two reservoirs at densities $\rho_a$ and $\rho_b$
is the same (up to boundary effects which give for large $L$ subleading contributions) as the probability of transporting the same current $j$ over a distance $L$ between two reservoirs at densities
$\rho_a$ and $\rho$ times the probability of transporting the current $j$
over a distance $L'$ between two reservoirs at densities $\rho$ and
$\rho_b$. One can then argue that choosing the optimal $\rho$ makes this probability maximum.
From (\ref{PPP1}) one gets the following additivity property of the large deviation function
\begin{equation}
F_{L+L'}\left(j,\rho_a,\rho_b\right) =
\max_\rho \left[
F_{L}\left(j,\rho_a,\rho\right) +
F_{L'}\left(j,\rho,\rho_b\right) \right] \; .
\label{FFF}
\end{equation}

Suppose that we consider a diffusive system for which we know the two functions $D(\rho)$ and $\sigma(\rho)$
defined   in (\ref{Ddef},\ref{sigmadef}).
If one accepts the additivity property (\ref{FFF}) of the large deviation
function, one can cut the system into more and more pieces so that

\begin{equation}
F_L\left(j,\rho_a,\rho_b\right) =
\min_{\rho_1,... \rho_{n-1}} \left\{\sum_{i=0}^{n-1}
 F_{L/n}  (j, \rho_i,\rho_{i+1})
\right\}
\label{FFFk}
\end{equation}
where $\rho_0=\rho_a$ and $\rho_n=\rho_b$.

For large $n$, the optimal choice of the $\rho_i$'s is such that the differences $\rho_i - \rho_{i+1}$ are small. For a current $j$ of order $1/L$, and for $\rho_i - \rho_{i+1}$ small,
one can replace  (\ref{Ddef},\ref{sigmadef})
$F_{l} $ for $l=L/k$
by 
\begin{equation}
F_l (j, \rho_i,\rho_{i+1}) \simeq { [ j - {D(\rho_i) (\rho_i-\rho_{i+1}) \over l}  ]^2 \over 2 {\sigma(\rho_i) \over l}}
\end{equation}
and by taking the limit $n \to \infty$ (keeping $l=L/n$ large for (\ref{Ddef},\ref{sigmadef}) to be still valid)
 one gets \cite{BD1}
\begin{equation}
F_{L}\left(j,\rho_a,\rho_b\right) = {1 \over L}\max_{\rho(x)} \left[ \int_0^1 {[jL +  D(\rho) \rho' ]^2 \over 2 \sigma(\rho) } dx \right]
\label{general}
\end{equation}
where the optimal profile $\rho(x)$ (for large $n$, the optimal $\rho_i$
in (\ref{FFFk}))   given by $\rho_i=\rho(i/n)$ should satisfy $\rho(0)= \rho_a$ and $\rho(1)= \rho_b$.

One can show \cite{BD1,BD4} that the optimal profile in (\ref{general})
(when $\rho_a \neq \rho_b$ and the deviation of current $j$ is small enough for this optimal profile to be still monotone)
is given by 
\begin{equation}
 \rho'(x)^2 ={(Lj)^2 (1+2 K \sigma(\rho(x)) \over D^2(\rho(x)) }
\label{rhooptimal}
\end{equation}
where the constant $K$ is adjusted to insure that $\rho(0)=\rho_a$ and $\rho(1)=\rho_b$.
Replacing $\rho(x)$ by (\ref{rhooptimal}) in (\ref{general}) leads to \begin{equation}
F_L(j,\rho_a,\rho_b)=  j \int_{\rho_b}^{\rho_a}  \left[ {1+ K \sigma(\rho) \over [1+2 K \sigma(\rho)]^{1/2}} -1 \right] {D(\rho) \over \sigma(\rho) }
\; d \rho
\label{LDJ}
\end{equation}
where the constant $K$ is fixed from (\ref{rhooptimal}) by the boundary
conditions  ($\rho(0)=\rho_a$ and $\rho(1)=\rho_b$)
\begin{equation}
Lj= \int_{\rho_b}^{\rho_a}  { D(\rho) \over [1+2 K \sigma(\rho)]^{1/2}}
 \; d \rho \  .
  \label{LDJ_current}
 \end{equation}
Expressions (\ref{LDJ},\ref{LDJ_current}) give therefore $F_L(j)$ in a parametric form.

The optimal profile (\ref{general}) remains unchanged when $j \to -j$ (simply  the sign of   $[1+2 K \sigma(\rho)]^{1/2}$ is changed)  in (\ref{LDJ},\ref{LDJ_current}) and one gets that
(\ref{fluctuation-dissipation-ter})
\begin{equation}
 F_L(j) - F_L(-j)= - 2 j \int_{\rho_b}^{\rho_a} {D(\rho) \over \sigma(\rho) } \;  d \rho
=-j(\log z_a - \log z_b) \  .
\label{GC1}
\end{equation}
Thus the expression (\ref{LDJ},\ref{LDJ_current}) does satisfy the
fluctuation theorem (\ref{fluctuation-theorem-3}).

From (\ref{LDJ},\ref{LDJ_current}) one can calculate $\mu(\lambda)$ by
(\ref{legendre1}) and one gets \cite{BD1,BD4} a parametric form
\begin{equation}
\mu (\lambda,\rho_a,\rho_b) =
-   {K  \over L} \left[ \int_{\rho_b}^{\rho_a}  { D(\rho) \ d \rho
\over \sqrt{ 1 + 2 K \sigma(\rho)}} \right]^2  \, ,
\label{mu}
\end{equation}
with $K = K(\lambda,\rho_a,\rho_b)$ is the solution of
\begin{equation}
\lambda=
\int_{\rho_b}^{\rho_a} d \rho { D(\rho) \over
\sigma(\rho)}\left[{1
\over \sqrt{ 1 + 2 K \sigma(\rho)}} -1 \right] \, .
\label{lambda}
\end{equation}
One can then get \cite{BD1}, by eliminating $K$ (perturbatively in $\lambda$) the expansion of $\mu(\lambda)$ in powers of $\lambda$ and therefore   the cumulants 
(\ref{cum})
 in the long time limit
for arbitrary $\rho_a$ and $\rho_b$
\begin{equation}
{\langle Q_t \rangle \over t} = {1 \over L} I_1,
\quad
 {\langle Q_t^2 \rangle - \langle Q_t
\rangle^2
\over t} = {1 \over L} {I_2\over I_1}, \quad
{\langle Q_t^3 \rangle_{ c} \over t}=  {1 \over L} { 3 (I_3 I_1 -
I_2^2) \over I_1^3} , \quad
{\langle Q_t^4 \rangle_{ c} \over t}=  {1 \over L} { 3 ( 5 I_4 I_1^2  - 14
I_1 I_2 I_3 +  9 I_2^3  ) \over I_1^5}
\label{cum1}
\end{equation}
where
\begin{equation}
I_n = \int_{\rho_b}^{\rho_a} D(\rho)\  \sigma(\rho)^{n-1} \  d \rho \; .
\end{equation}
Using the fact (\ref{sigmaSSEP}) that for the SSEP $D(\rho)=1$ and $\sigma(\rho) = 2
\rho(1- \rho)$,  one can recover \cite{BD1} from
(\ref{mu},\ref{lambda},\ref{cum1})  the above expressions (\ref{q1c1}-\ref{q4c1},\ref{Fomega}).
The validity of (\ref{LDJ},\ref{LDJ_current}) has also been checked for weakly interacting lattice gases \cite{WR}.
\\ \ \\ 
We have seen in \ref{mft} that
the macroscopic fluctuation theory gives the probability (\ref{eq: dev exp})
of  arbitrary (rescaled) density and current profiles $\widehat{\rho},\widehat{j}$. Therefore to observe (\ref{F(j)}) an average current $j$ over a long time $t$ one should have
\begin{equation}
F(j) = \lim_{t \to \infty}{1 \over t}  L  \min_{ \widehat{\rho}(x,\tau), \widehat{j}(x,\tau)}
\int_{0}^{t/L^2}
d \tau \int_0^1 dx { \left[\widehat{j}(x,\tau) + D({\widehat{\rho}(x,\tau}))
{\partial {\widehat{\rho}(x,\tau)} \over \partial x}\right]^2 \over 2
\sigma(\widehat{\rho}(x,\tau))}
\label{general-bis}
\end{equation}
with the constraint that
\begin{equation}
jL = {L^2 \over t} \int_0^{t/L^2} \widehat{j}(x,\tau) d \tau \  .
\end{equation}
Comparing (\ref{general}) and (\ref{general-bis}) we see that the two expressions coincide when the optimal $\widehat{\rho},\widehat{j}$ in (\ref{general-bis})  are independent of the time $\tau$.
 Therefore the additivity principle gives the large deviation function $F(j)$ of the current  only when the optimal profile in 
 (\ref{general-bis}) is time independent. 

Bertini et al  \cite{BDGJL5,BDGJL6} pointed out that it can  happen, for some $\sigma(\rho)$
and $D(\rho)$, that the expression 
  (\ref{LDJ},\ref{LDJ_current})
of the large deviation function $F(j)$
 is  non-convex  (as it should) 
 in which case the expression (\ref{general})  is no longer valid (and
the prediction  (\ref{general}) 
becomes \cite{ BDGJL5,BDGJL6}  simply an upper bound  of $F(j)$).
This is because the optimal profile in (\ref{general-bis}) is no longer
constant in time. 
 When this optimal profile   is time dependent, 
one has to solve a much harder optimisation  problem \cite{BD2,BD4} than  (\ref{general}).

Restrictions on $\sigma(\rho)$ and $D(\rho)$ for
(\ref{general}) to be valid have been given in \cite{BDGJL6} and there are
cases , such as the WASEP on a ring  \cite{BD2,BD4}, for which by varying $\lambda$ or
the asymmetry one can observe a phase transition between a phase where
the optimal profile in (\ref{general-bis}) is constant in time and a
phase where it becomes time dependent.

\section{The matrix approach for the  asymmetric exclusion process}
\label{matrix-tasep}

The matrix ansatz of section  \ref{matrix-ssep} (which gives the weights of the
microscopic configurations in the steady state) has been generalized
to describe the steady state of several other systems 
\cite{DJLS,DEM,Sandow,DEmal,StinchS,Dmal,ER,HSP,MS,DLS,Mal,KS,Speer,DGLS,Sasamoto,ADR,BECE,HH,IPR,BFMM,Jafarpour,DST1,DST2}, with of course modified algebraic rules for the matrices the vectors $\langle W|$ and $|V \rangle$.

For example for the asymmetric exclusion process (ASEP), for which the
definition is the same as the SSEP  except that particles
jump at rate $1$ to their right
and at rate $q \neq1$ to their left it the target site is empty (see figure \ref{asep}),
one can show \cite{DEHP,Sandow,Sasamoto,BECE} that the weights are still given
by (\ref{matrix}) with the algebra (\ref{algebra}) replaced by
\begin{align}\label{eqn:commutation_DE_ASEP}
  DE-qED & = D+E \\
  \langle W| (\alpha E-\gamma D) &= \langle W|\\
  (\beta D-\delta E)|V\rangle &= |V\rangle \  .
\end{align}
One should notice that for the ASEP, the direct approach of calculating,
as in section \ref{ssep-sec}, the steady state properties by writing the time
evolution leads nowhere. Indeed (\ref{evolution}) becomes
\begin{align}
{d \langle \tau_1 \rangle \over dt } = & \alpha - (\alpha + \gamma + 1 )
\langle \tau_1 \rangle + q \langle \tau_2 \rangle +(1-q) \langle \tau_1 \tau_2 \rangle
\nonumber \\
{d \langle \tau_i \rangle \over dt } = &
\langle \tau_{i-1} \rangle -(1+q)  \langle \tau_i \rangle +  q \langle
\tau_{i+1} \rangle      -(1-q) (\langle \tau_{i-1} \tau_i \rangle
-\langle \tau_{i} \tau_{i+1} \rangle)
                   \label{evolution1}                                     \\
{d \langle \tau_L \rangle \over dt } = &
\langle \tau_{L-1} \rangle -(q+ \beta + \delta)  \langle \tau_L \rangle + \delta- (1-q) \langle \tau_{L-1} \tau_L \rangle
   \nonumber                 \end{align}
and the equations which determine  the one-point functions are no longer
closed.
Therefore all the correlation functions have to be determined at the same time
and this is what the matrix ansatz (\ref{matrix}) does.
Alternative combinatorial methods to calculate the steady state weights
of exclusion processes with open boundary conditions have been obtained
in \cite{BE,Duchi}.

The large deviation function ${\cal F}$ of the density defined by
(\ref{LDdef}) has been calculated for the ASEP \cite{DLS3,DLS4,ED} by  an
extension of the approach of sections \ref{add-ssep} and
\ref{ld-dens-ssep} (see section \ref{add-tasep}).
\section{The phase diagram of the totally asymmetric exclusion process}
\label{phas-diag}
The last three sections \ref{phas-diag}-\ref{brown-excurs} present, as examples,
three results which can be obtained rather easily for  the totally asymmetric case (TASEP) i.e. for $q=0$
(in the particular case where particles are injected only at the left boundary and removed only at the right boundary i.e. when the input rates $\gamma=\delta=0$). In this case the algebra (\ref{eqn:commutation_DE_ASEP}) becomes
\begin{align}\label{eqn:commutation_DE_TASEP}
  DE & = D+E \\
\label{eqn:commutation_DE_TASEP-1}
  \langle W| \alpha E &= \langle W|\\
  \beta D|V\rangle &= |V\rangle
\label{eqn:commutation_DE_TASEP-2}
\end{align}
As for the SSEP the average current $\langle J \rangle$ is still given in terms of the vectors $\langle W|$, $ V\rangle $ and of the matrices $D$ and $E$ by
\begin{equation}
\label{current-ASEP}
\langle J \rangle=
 { \langle W | (D+E)^{L-1} | V
\rangle \over \langle W | (D+E)^{L} | V
\rangle } \ .
\end{equation}
However as the algebraic rules have changed, the expression of the current is different for the SSEP and the ASEP.
From the relation $DE=D+E$ it is easy to prove  by recurrence that
\begin{equation*}
   DF(E) = F(1)D + E \:\frac{F(E)-F(1)}{E-1}
\end{equation*}
for any polynomial $F(E)$ and
\begin{equation*}
  (D+E)^N = \sum_{p=1}^N
  \frac{p(2N-1-p)!}{N!(N-p)!}\big(E^p+E^{p-1}D+\ldots+D^p\big)\:.
\end{equation*}
Using the fact that
\begin{equation*}
 \frac{\langle W|E^mD^n|V\rangle}
       {\langle W|       V\rangle}
= \frac{1}{\alpha^m} \frac{1}{\beta^n}\:,
\end{equation*}
one gets \cite{DEHP}
\begin{equation} \label{eqn:DE_TASEP}
  \frac{\langle W|(D+E)^N|V\rangle}
      {\langle W|        V\rangle}
  \;=\;
  \sum_{p=1}^N\;
  \frac{p(2N-1-p)!}{N!(N-p)!} \;
  \frac{\dfrac{1}{\alpha^{p+1}}-\dfrac{1}{\beta^{p+1}}}
       {\dfrac{1}{\alpha      }-\dfrac{1}{\beta}} \:.
\end{equation}
For large $N$ this sum is dominated either by $p\sim 1$, or $p\sim N$ depending one the values of $\alpha$ and $\beta$ and one obtains
\begin{equation}
\frac{\langle W|(D+E)^N|V\rangle}
      {\langle W|        V\rangle}
\sim \left\{\begin{aligned}
   & 4^N & \quad\text{if}\quad \alpha > \frac{1}{2} \text{ and }
                              \beta  > \frac{1}{2} \\
   &[\beta(1-\beta)]^{-N}
        & \quad\text{if}\quad \beta  < \alpha \text{ and }
                              \beta   < \frac{1}{2} \\
   &[\alpha(1-\alpha)]^{-N}
        & \quad\text{if}\quad \beta  > \alpha \text{ and }
                              \alpha  < \frac{1}{2}  \:.\\
  \end{aligned}
\right.
\label{norm1}
\end{equation}
This leads to three different expressions of
the current (\ref{current-ASEP}) for  large $L$ corresponding to the three different phases:
\begin{itemize}
\item
the low density phase ($\beta  > \alpha $ and
                              $\alpha  < \frac{1}{2}$) where
$\langle J \rangle = \alpha(1-\alpha)$
\item
the high density phase ($\alpha  > \beta $ and
                              $\beta   < \frac{1}{2}$) where
$\langle J \rangle = \beta (1-\beta )$
\item
the maximal current phase ($\alpha > \frac{1}{2}$  and
                              $\beta  > \frac{1}{2}$)
where $\langle J \rangle = {1 \over 4}$
\end{itemize}
which is the
exact  phase diagram of the TASEP
\cite{K,DDM,DEHP,SD}.
The existence of phase transitions \cite{KSKS,PS,KLMST,Evans}      in these driven lattice gases is one of
the  striking properties of   non-equilibrium steady states, as it is
well known that one dimensional systems at equilibrium with short range
interactions  cannot exhibit phase transitions.

\section{ Additivity and large deviation function for the TASEP}
\label{add-tasep}
Let us now see how the additivity relation (\ref{fermeture})  can be generalized for the TASEP in order to obtain the large deviation functional of the density.
For the algebra (\ref{eqn:commutation_DE_TASEP}-\ref{eqn:commutation_DE_TASEP-2}), if one inroduces the following eigenvectors
\begin{equation}
\label{eigen-TASEP} \langle \rho |E  = 
{1 \over \rho} \langle \rho |   \ \ \ \ ;   \ \ \ \ D | \rho \rangle = {1 \over 1-\rho}  | \rho \rangle 
\end{equation}
it is clear that
\begin{equation}
\langle W | = \langle \rho_a |  \ \ \ \ ;   \ \ \ \ 
|V \rangle = | \rho_b \rangle 
\end{equation}
with $\rho_a= \alpha$ and $\rho_b=1- \beta$.
Note that in general $\langle \rho_a|\rho_b \rangle \neq 0$ even when $\rho_a \neq \rho_b$.
Now one can prove, as in (\ref{fermeture}),  that for $\rho_b < \rho_a$
 \begin{equation}
\label{eqn:rel_fermeture_TASEP} 
  \frac{\langle \rho_a|Y_1Y_2|\rho_b\rangle}
       {\langle \rho_a|       \rho_b\rangle}
  = \oint\displaylimits_{\rho_b<|\rho|<\rho_a} \frac{d\rho}{2i\pi}\:
    \frac{(\rho_a-\rho_b)}{(\rho_a-\rho)(\rho-\rho_b)} \:
    \frac{\langle \rho_a|Y_1|\rho\rangle}
         {\langle \rho_a|    \rho\rangle} \:
    \frac{\langle \rho|Y_2|\rho_b\rangle}
         {\langle \rho|    \rho_b\rangle} 
\end{equation}
where $Y_1$ and $Y_2$  are arbitrary polynomials in $D$ and $E$. 
\\ \ \\
{\bf Proof of (\ref{eqn:rel_fermeture_TASEP}) :}
Any polynomial $Y$ of the operators $D$ and $E$ can be written, using
$DE=D+E$, as
\begin{equation}
Y = \sum_{n,n'} a_{n,n'} E^n D^{n'}
\end{equation}
by pushing all the $D$'s to the right and all the $E$'s to the left.
Therefore to prove (\ref{eqn:rel_fermeture_TASEP}) 
 it is sufficient to do it for
$Y_1$ and $Y_2$ of the form
\begin{equation}
Y_1 = E^{n_1} D^{n'_1} \ \ \ \ , 
\ \ \ Y_2 = E^{n_2} D^{n'_2}  \ .
\end{equation}
If $n'_1=0$ or $n_2=0$, the identity (\ref{eqn:rel_fermeture_TASEP}) is
easy to check. Then one can prove it by recursion on $n'_1+n_2$: if $Y_1=
Z_1 D$ and $Y_2 = E Z_2$, and one assumes that the identity
(\ref{eqn:rel_fermeture_TASEP}) is valid for
$Z_1D Z_2$ and $Z_1 E Z_2$, the l.h.s. of (\ref{eqn:rel_fermeture_TASEP})  
can be written as (\ref{eqn:commutation_DE_TASEP})
 \begin{equation}
\nonumber
 \frac{\langle \rho_a|Z_1DE Z_2|\rho_b\rangle} {\langle \rho_a|       \rho_b\rangle}
= \frac{\langle \rho_a|Z_1(D+E) Z_2|\rho_b\rangle} {\langle \rho_a|       \rho_b\rangle}
  = \oint\displaylimits_{\rho_b<|\rho|<\rho_a} \frac{d\rho}{2i\pi}\:
    \frac{(\rho_a-\rho_b)}{(\rho_a-\rho)(\rho-\rho_b)} \left[{1 \over \rho} + {1
\over  1- \rho} \right] \:
    \frac{\langle \rho_a|Z_1 |\rho\rangle}
         {\langle \rho_a|    \rho\rangle} \:
    \frac{\langle \rho|Z_2|\rho_b\rangle}
         {\langle \rho|    \rho_b\rangle}
\end{equation}
and  simply beacuse $1/\rho+ 1/(1-\rho)= 1/(\rho(1-\rho))$  this becomes
\begin{equation}
\nonumber
\oint\displaylimits_{\rho_b<|\rho|<\rho_a} \frac{d\rho}{2i\pi}\:
    \frac{(\rho_a-\rho_b)}{(\rho_a-\rho)(\rho-\rho_b)} {1 \over
\rho
( 1- \rho)}\:
    \frac{\langle \rho_a|Z_1 |\rho\rangle}
         {\langle \rho_a|    \rho\rangle} \:
    \frac{\langle \rho|Z_2|\rho_b\rangle}
         {\langle \rho|    \rho_b\rangle}
=
\oint\displaylimits_{\rho_b<|\rho|<\rho_a} \frac{d\rho}{2i\pi}\:
    \frac{(\rho_a-\rho_b)}{(\rho_a-\rho)(\rho-\rho_b)} 
    \frac{\langle \rho_a|Z_1 D |\rho\rangle}
         {\langle \rho_a|    \rho\rangle} \:
    \frac{\langle \rho|E Z_2|\rho_b\rangle}
         {\langle \rho|    \rho_b\rangle}
\end{equation}
which is the r.h.s. of (\ref{eqn:rel_fermeture_TASEP}).
\ \\ \ \\
We are now going to see, as an example, how
 the large deviation function $\cal F$ of the density   can be derived for the TASEP from (\ref{eqn:rel_fermeture_TASEP})
when $\rho_a > \rho_b$. If one defines $K(\rho_a,\rho_b)$ by
\begin{equation}
\label{Krhoarhobdef}
K(\rho_a,\rho_b)= \lim_{L \to \infty}{1 \over L}\log {\langle \rho_a |(D+E)^L |\rho_b \rangle \over  \langle \rho_a |\rho_b \rangle }  
\end{equation}
one can easily check  from (\ref{norm1}) that for
$\rho_b(=1- \beta) < \rho_a (= \alpha)$ 
\begin{equation}
K(\rho_a,\rho_b) =
-\max_{\rho_b<\rho<\rho_a} \log(\rho(1-\rho))
\label{Krhoarhobdef-bis}
\end{equation}
Using a saddle point method in (\ref{eqn:rel_fermeture_TASEP})
 when $Y_1$ and $Y_2$ are sums of long products of $D$'s and $E$'s, one gets 
\begin{equation}
  \frac{\langle \rho_a|Y_1Y_2|\rho_b\rangle}
       {\langle \rho_a|       \rho_b\rangle}
\simeq  \min_{\rho_b \leq F \leq \rho_a}
  \frac{\langle \rho_a|Y_1|F\rangle}
       {\langle \rho_a|       F\rangle}
  \frac{\langle F|Y_2|\rho_b\rangle}
    {\langle F|       \rho_b\rangle}
\label{cut}   
\end{equation}
(Note that in applying the saddle point method, one needs to find the maximum $F$ over the circular integration contour. This maximum   is at the same time a {\it minimum} when $F$ varies along the real axis). Then as for a system of large size $L$ 
(\ref{Krhoarhobdef},\ref{Krhoarhobdef-bis}) one has
\begin{equation}
  \frac{\langle \rho_a|Y|\rho_b\rangle}
       {\langle \rho_a|(D+E)^L |      \rho_b\rangle}
\sim 
e^{- K(\rho_a\rho_b) \: L} \ 
 \frac{\langle \rho_a|Y|\rho_b\rangle}
       {\langle \rho_a|       \rho_b\rangle}
\end{equation}
One can of course repeat  (\ref{cut}) several times to relate  a large system of size $L$ to its subsystems (as long as these subsystems are large). Therefore one gets
\begin{equation}
  \frac{\langle \rho_a|Y_1Y_2 ... Y_k|\rho_b\rangle}
       {\langle \rho_a| (D+L)^L    |  \rho_b\rangle}
\sim
e^{- K(\rho_a\rho_b) \; L}
\min_{\rho_a \geq F_1 \geq F_2 ...\geq F_{n-1} > \rho_b}
\prod_{i=1}^n \left[
  \frac{\langle F_{i-1}|Y_i|F_i\rangle}
       {\langle F_{i-1}|  (D+E)^l |     F_i\rangle}
e^{ K(F_{i-1},F_i) \; l  }\right]
\label{cut-bis}   
\end{equation}
where $F_0= \rho_a$, $F_n= \rho_b$ and $l=L/n$.
If $Y_i$ is the sum of the matrix elements of  all configurations of $l$
sites with $ l \rho_i$ particles one gets for the large deviation function ${\cal F}_n$ defined  in (\ref{finite})
\begin{equation}
\label{Fn-bis}
{\cal F}_n(\rho_1, \rho_2, ...\rho_n|\rho_a,\rho_b)
=  K(\rho_a,\rho_b)+
 \max_{\rho_b=F_0 < F_1  ..  <F_i < ..< F_n=\rho_a}
{1 \over n}\sum_{i=1}^n{\cal F}_1(\rho_i|F_{i-1},F_i)  - K(F_{i-1},F_i) \ .
\end{equation}
For large $n$, almost all the differences $F_{i-1} - F_i$ are small, so that 
$${\cal F}_1(\rho_i|F_{i-1},F_i) 
\simeq {\cal F}_1(\rho_i|F_{i},F_i) = \rho_i \log{\rho_i \over F_i}
+(1- \rho_i) \log{1-\rho_i \over 1-F_i} \equiv B(\rho_i,F_i)$$ since when
the two densities $F_{i-1}$ and $F_i$  are equal, the steady state measure is Bernoulli
and this leads to 
\begin{equation}
  {\cal F} (\{\rho(x)\}) = - \max_{\rho_b<r<\rho_a}  \Big[
    \log\big(r(1-r)\big)\Big] + \sup_F \int_0^1dx\: \left[
    B\big(\rho(x),F(x)\big)+\log \Big( F(x)\big(1-F(x)\big) \Big) \right]\:,
\end{equation}
which is the expression of the large deviation function of the density of the TASEP (and also of the ASEP \cite{DLS3,DLS4}) for $\rho_a > \rho_b$.

For $\rho_a> \rho_b$,
 one can also obtain 
\cite{DLS3,DLS4,ED}
this large deviation function, 
starting from a relation similar to 
(\ref{eqn:rel_fermeture_TASEP}) obtained by deforming the circular contour to insure that $\rho_b$ remains inside and $\rho_a$ outside.
One can note that when $Y_1$ and $Y_2$ are polynomials in $D$ and $E$, all the matrix elements in (\ref{eqn:rel_fermeture_TASEP}) are rational functions of $\rho_a$ and $\rho_b$ which can be easily anaytically continued from the case $\rho_a> \rho_b$ to the case $\rho_a < \rho_b$. The result is
\cite{DLS3,DLS4}
\begin{equation}
\label{func-tasep}
  \begin{aligned}
 {\cal F}(\{\rho(x)\}) = \inf_{0<y<1}  
   &\Big[\int_0^y dx\:\left( B(\rho(x),\rho_a)+
    \log\frac{\rho_a(1-\rho_a)}
             {\langle J \rangle} \right)\\
   &+\int_y^1 dx\:\left( B(\rho(x),\rho_b)+
    \log\frac{\rho_b(1-\rho_b)}
             {\langle J \rangle} \right)\Big]
  \end{aligned}\:.
\end{equation}
For the TASEP one knows \cite{DEHP,SD} that  along the line
$\rho_a=1-\rho_b <{1 \over 2}$  there is a first order phase transition
line. Along this line $\langle J \rangle = \rho_a(1-\rho_a) = \rho_b (1-
\rho_b)$ and  the typical configurations  $\rho_z(x)$ are schocks \cite{DLS,DGLS,SA} located at
arbitrary positions  $z$ beween a region of density $\rho_a$ and a region of density $1 - \rho_a$.
\begin{equation}
\rho_z(x) = \begin{cases} \rho_a 
\ \ \ \  \ \ \ \ {\rm for} \ \   0 < x < z \cr
\\
 1 - \rho_a 
\ \ \ \  \ {\rm for} \ \    z  < x < 1 
\end{cases}
\end{equation}
For all these profiles $\rho_z(x)$, the functional 
(\ref{func-tasep}) vanishes. It is also easy to check that ${\cal
F}(\rho(x)) > 0$   for a profile
of the form
\begin{equation}
\rho(x) =\alpha \rho_z(x) + (1 - \alpha) \rho_{z'}(x)
\end{equation}
and this shows that $\cal F$ is non-convex. Therefore in contrast to equilibrium systems, the functional  ${\cal F}(\rho(x))$ may be non-convex in non-equilibrium steady states.

\section{Correlation functions in the TASEP and Brownian excursions}
\label{brown-excurs}

In this last example, we will see that the  fluctuations of the density
 are  non-Gaussian in the maximal current phase of the TASEP.
In this  maximal current phase ($\alpha > \frac{1}{2}$  and
                              $\beta  > \frac{1}{2}$)
one can show \cite{DEL}, using the matrix ansatz, that
the correlation  function of the  occupations of $k$ sites at positions
$i_1 = L x_1, i_2=Lx_2,... i_k=L x_k$ with
$ x_1 <  x_2 <... <  x_k$
are given by
\begin{equation} \label{eqn:brown_excurs}
 \left\langle
  \left(\tau_{L x_1}-\frac{1}{2}\right) \ldots
  \left(\tau_{L x_k}-\frac{1}{2}\right)
 \right\rangle =
  \frac{1}{2^k} \frac{1}{L^{k/2}}
  \frac{d^k}{dx_1\ldots dx_k}
  \Big\langle y(x_1) \ldots y(x_k) \Big\rangle \:,
\end{equation}
where  $y(x)$ is a Brownian excursion between $0$ and $1$
(a Brownian excursion is a Brownian path constrained to $y(x) > 0$ for $0 < x <1$
with the boundaries $y(0)=y(1)$).
The probability
$P\big(y_1\ldots y_k;x_1\ldots x_k\big)$ of finding the Brownian excursion at positions $y_1\ldots y_k$
for  $0<x_1<\ldots<x_k<1$ is
\[
  P\big(y_1\ldots y_k;x_1\ldots x_k\big) = \frac{
  h_{x_1}(y_1)\:g_{x_2-x_1}(y_1,y_2)\ldots
  g_{x_{k}-x_{k-1}}(y_{k-1},y_k)\:h_{1-x_k}(y_k)}{\sqrt{\pi}}\:,
\]
where $h_x$ and $g_x$  are defined by
\[
  \left\{\begin{aligned}
    h_x(y) &=\frac{2y}{x^{3/2}} e^{-y^2/x}\\
    g_{x}(y,y')&=\frac{1}{\sqrt{\pi x}}\left(
                e^{-(y-y')^2/x}-e^{-(y+y')^2/x}\right)\:.\\
  \end{aligned}\right.
\]

One can derive easily (\ref{eqn:brown_excurs}) in the particular case
 $\alpha=\beta=1$ using a  representation of (\ref{eqn:commutation_DE_TASEP})
which consists of two infinite dimensional bidiagonal matrices
\begin{align*}
D&=\sum_{n\geq 1} |n\rangle\langle n|+|n\rangle\langle n+1| =
  \begin{pmatrix}
   1  & 1      &0        & 0 & 0 & 0  & \cdots \\
   0  & 1      &1        & 0 & 0 & 0  & \cdots \\
   0  & 0      &1        & 1 & 0 & 0  & \cdots \\
   0  & 0      &0        & 1 & 1 & 0  & \cdots \\
      &        &         &   & \ddots & \ddots &  \\
  \end{pmatrix} \\
E&=\sum_{n\geq 1} |n\rangle\langle n+1|+|n\rangle\langle n| =
  \begin{pmatrix}
   1  & 0      &0        & 0 & 0 & 0  & \cdots \\
   1  & 1      &0        & 0 & 0 & 0  & \cdots \\
   0  & 1      &1        & 0 & 0 & 0  & \cdots \\
   0  & 0      &1        & 1 & 0 & 0  & \cdots \\
      &        &         &  \ddots & \ddots &  &  \\
  \end{pmatrix} \\
\end{align*}
with
\begin{align*}
\langle W| &= \langle 1| = (1,0,0\ldots) \\
\langle V| &= \langle 1| = (1,0,0\ldots) \:.
\end{align*}
With this representation one can write  $\langle W|(D+E)^L|V\rangle$ as a
sum over a set
$\mathcal M_L$
of one dimensional random walks $w$
of  $L$  steps
which remain positive. Each walk  $w$  is defined by
a sequence  $\big(n_i(w)\big)$ of $L-1$ heights  $(n_i(w)\geq 1)$  (with
at the boundaries $n_0(w)=n_L(w)=1$ and the constraint
$|n_{i+1}-n_i|\leq 1$)~:
\[
  \langle W|(D+E)^L|V\rangle = \sum_{w\in\mathcal M_L} \Omega (w)\:,
\]
where
\[
  \Omega (w) = \prod_{i=1}^L v\big(n_{i-1},n_i\big)
  \quad\text{with}\quad
  v\big(n,n'\big) = \left\{\begin{aligned}
                      2 & \quad\text{if }~ |n-n'|=0 \\
                      1 & \quad\text{if }~ |n-n'|=1\:. \\
                    \end{aligned}\right.
\]
One has $v\big(n,n'\big)=\langle n|D+E|n'\rangle$  since
$D+E$ has a tridiagonal form
\[
  D+E=
  \begin{pmatrix}
   2  & 1      &        & (0) \\
   1  & \ddots & \ddots &     \\
      & \ddots & \ddots & 1   \\
  (0) &        &    1   & 2   \\
  \end{pmatrix} \;.\\
\]
Then from the matrix expression one gets
$\langle\tau_i\rangle$ et $\langle\tau_i \tau_j\rangle$~:

\begin{equation} \label{eqn:discrete_brown_excurs}
 \left\langle
  \left(\tau_{i_1}-\frac{1}{2}\right) \ldots
  \left(\tau_{i_k}-\frac{1}{2}\right)
 \right\rangle =
 {1 \over 2^k} \sum_w \nu(w)
  \big(n_{i_1}-n_{i_1-1}\big)\ldots \big(n_{i_k}-n_{i_k-1}\big)\:,
\end{equation}
where  $\nu(w)$ is the probability of the walk $w$ induced  by the
weights  $\Omega$~:
\[
  \nu(w)=\frac{\Omega(w)}{\sum_w'\Omega(w')}\:.
\]
The expression (\ref{eqn:discrete_brown_excurs}) is the discrete version
of
(\ref{eqn:brown_excurs}).  The result (\ref{eqn:brown_excurs}) can be extended
\cite{DEL}
 to arbitrary values of $\alpha$ and  $\beta$ in the maximal current
phase (i.e. for $\alpha > 1/2$ and $\beta > 1/2$).

From this link between the density fluctuations and Brownian excursions,
one can show
that, for a TASEP of $L$ sites,  the number $N$ of particles
beween sites $Lx_1$ and  $Lx_2$,  has non-Gaussian fluctuations in the maximal current phase:
if one defines the reduced density
\begin{equation}
   \mu = \frac{N-L(x_2-x_1)/2}{\sqrt{L}}\:.
\label{mu1}
\end{equation}
one can show \cite{DEL} that for large $L$
\begin{equation}\label{eqn:fluct_non_gauss}
  P(\mu) = \int_0^\infty dy_1 \int_0^\infty dy_2 \:
  \frac{1}{\sqrt{2\pi(x_2-x_1)}}\:
  \exp\left(-\frac{(2\mu+y_1-y_2)^2}{x_2-x_1}\right)\:.
\end{equation}
This contrasts with the Gaussian fluctuations of the density
(\ref{pvbis}) at equilibrium.
According to numerical simulations \cite{DEL} the distribution
(\ref{mu1},\ref{eqn:fluct_non_gauss})
(properly rescaled) of the fluctuations of the density  remains valid for more 
general driven systems  in their maximal current phase.
Of course proving it in a more general case is an interesting open
question.

\section{Conclusion}
\label{conclusion}

These lectures  have presented a number of recent results concerning the
fluctuations and the large deviation functions of the density or of the
current in non-equilibrium steady states.
\\ \ \\
For general diffusive systems, the macroscopic fluctuation theory
\cite{BDGJL1,BDGJL2,BDGJL3,BDGJL4,BDGJL5,BDGJL6} discussed in  section \ref{mft} gives a framework to calculate the large deviation of the density  ${\cal
F}(\rho(x))$ leading to  equations (\ref{hamilton-jacobi},\ref{dyn},\ref{j-bertini}) which are in general difficult to solve. 
One can however check that the expressions (\ref{F2},\ref{F3},\ref{F5}) obtained
\cite{DLS1,DLS2} in sections \ref{matrix-ssep}-\ref{ld-dens-ssep} by the matrix ansatz do solve these equations.
So far the large deviation functional is only known for very few examples
 \cite{DLS1,DLS2,BGL,ED}. There remains a lot to be done to understand the general
properties of the functional ${\cal F}(\rho(x))$. For example, with
Thierry Bodineau we tried, so far without success, to calculate ${\cal F}(\rho(x)) $ (\ref{hamilton-jacobi},\ref{dyn},\ref{j-bertini})
for general $D$ and $\sigma$ in powers of $\rho_a-\rho_b$. Also  for
the SSEP, we did not succeed to obtain the large deviation functional of
the density ${\cal F}(\rho(x))$ for the $\lambda-$measure  defined in  (\ref{H(C)}). Other situations which would be interesting to consider are the cases of several species of particles, several conserved quantities \cite{Olla1,Olla2} or non-conserved quantities \cite{BJ}.
\\ \ \\
For driven diffusive systems, the situation is worse: so far ${\cal
F}(\rho(x))$ is only known for the ASEP \cite{DLS3,DLS4} and, to my
knowledge, there does not exist so far a general theory to calculate this
large deviation functional for  general driven diffusive systems \cite{BD3}. In contrast to equilibrium systems, for the ASEP, the large deviation functional may be non-convex (section \ref{add-tasep}) or  fluctuations of the density may be non-Gaussian (section \ref{brown-excurs}). 
\\ \ \\ \ \\
For current fluctuations in  diffusive systems, the additivity principle
gives explicit expressions
(section \ref{add})
of the large deviation function of the current for general diffusive
systems. In some cases, however, these expressions are not valid, when the profile to produce a deviation of current becomes time dependent. In such cases the calculation of the large deviation function $F(j)$ of the current is much harder \cite{BD4}.
So far the predictions
(\ref{general},\ref{LDJ},\ref{LDJ_current},\ref{mu}-\ref{cum1}) of the additivity principle remain to be checked in specific  examples: even for the SSEP, only the first four cumulants are known to agree with (\ref{cum1}) but a direct calculation of $F(j)$ for the SSEP is, to my knowledge, still missing.
It would be nice to see whether this could be done by a Bethe ansatz calculation for the SSEP with open boundaries  \cite{GE}.
It would also be useful to test the predictions  of the additivity principle on other diffusive systems
and to try to extend them to more complicated situations, in particular
when there are more than a single conserved quanity \cite{Olla1,Olla2}.
\\ \ \\ \ 
 Concerning the  fluctuations or the large deviations of the current of
driven diffusive systems, there has been lots of progress
 over the last ten years \cite{J,PS1,PS2,Saamoto,H5,BFPS}. On the infinite line
exact results for the TASEP and the PNG (polynuclear growth model) lead
to a universal distribution of current characteristic of the KPZ universality
class. On the ring too, Bethe ansatz calculations
\cite{GS,Schutz,DL,Der,DA,kim,Pri,BPS,OG1,RS,PP,OG2},
allow  to calculate the large deviation function of the current which
exhibits a universal shape in the scaling regime. For driven diffusive
systems, however, there is not yet a general approach allowing to calculate the large
deviation function or the fluctuations of the current  for all
geometries, including finite systems with open boundary conditions
\cite{GE}. Of course it would be nice to extend the macroscopic fluctuation theory 
to get a framework allowing to calculate both the  large deviation
functions of the current and of the density for general
 driven diffusive systems. 
\\ \ \\ \  \\ \ \\

I would like to thank the Newton Institute Cambridge for its hospitality and the organizers of the  workshop "Non-Equilibrium Dynamics of Interacting Particle
Systems"   for inviting me to deliver the lectures on which these notes are based.
These notes attempt to give  a summary of results obtained in a  series of collaborations over the last 15 years. I would like to thank 
C. Appert, T. Bodineau, M. Clincy, B. Dou\c{c}ot, C. Enaud, M.R. Evans,
S. Goldstein, V. Hakim, S.A. Janowsky, C. Landim, J.L. Lebowitz,  K.
Mallick, D. Mukamel, S. Olla, V. Pasquier, P.E. Roche, E.R. Speer with
whom I had the pleasure and the privilege to work on these topics.

\end{document}